\documentclass{aa}  
\usepackage{bm}
\usepackage{color}
\usepackage{float}
\usepackage{graphicx,amsmath}
\usepackage[varg]{txfonts}
\usepackage{natbib}
\DeclareGraphicsExtensions{.pdf,.png,.jpg}

\def\wig#1{\mathrel{\hbox{\hbox to 0pt{%
          \lower.6ex\hbox{$\sim$}\hss}\raise.4ex\hbox{$#1$}}}}
\def\acc{\alpha_{\rm acc}}
\def\Dr{\alpha_{{\rm D}r}}
\def\Dz{\alpha_{{\rm D}z}}

\def\rhog{\rho_{\rm g}}
\def\rhod{\rho_{\rm d}}

\def\mdotp{\dot{M}_{\rm p}}
\def\mdotg{\dot{M}_{\rm g}}

\def\hg{H_{\rm g}}
\def\hp{H_{\rm p}}
\def\hd{H_{\rm d}}
\def\hdg{h_{\rm d/g}}
\def\h0{h_{\rm d/g,0}}
\def\hpg{h_{\rm p/g}}
\def\fpg{F_{\rm p/g}}
\def\vr{\varv_r}  
\def\st{\tau_{\rm s,d}}
\def\stpeb{\tau_{\rm s,p}}

\def\sigd{\Sigma_{\rm d}}
\def\sigg{\Sigma_{\rm g}}
\def\vk{\varv_{\rm K}}

\def\ga{\ \lower 3pt\hbox{${\buildrel > \over \sim}$}\ }
\def\la{\ \lower 3pt\hbox{${\buildrel < \over \sim}$}\ }

\usepackage[normalem]{ulem}

\graphicspath{{figs/}{./}}
\begin{document} 
   \title{Planetesimal formation around the snow line:
   I. Monte Carlo simulations of silicate dust pile-up in a turbulent disk}
   \author{Shigeru Ida
          \inst{1}
          \and
          Tristan Guillot
          \inst{2}
          \and
          Ryuki Hyodo
          \inst{3}
           \and
          Satoshi Okuzumi
          \inst{4}
           \and
          Andrew N. Youdin
          \inst{5}
           }
   \institute{Earth-Life Science Institute, Tokyo Institute of Technology, Meguro-ku, Tokyo 152-8550, Japan \\
              \email{ida@elsi.jp}
         \and
             Laboratoire J.-L.\ Lagrange, Universit\'e C\^ote d'Azur, Observatoire de la
  C\^ote d'Azur, CNRS, F-06304 Nice, France \\
         \and
             ISAS/JAXA, Sagamihara, Kanagawa, Japan \\ 
         \and
             Department of Eartha and Planetary Sciences, Tokyo Institute of Technology, Meguro-ku, Tokyo 152-8551, Japan \\
              \and
          Steward Observatory/The Lunar and Planetary Laboratory, University of Arizona, Tucson, Arizona 85721, USA.  
           }
   \date{DRAFT:  \today}
  \abstract{   
The formation of rocky planetesimals is a long-standing problem in planet formation theory. One of the possibilities is that it results from gravitational instability as a result of pile-up of small silicate dust particles released from sublimating icy pebbles that pass the snow line.  
} 
{
We want to understand and quantify the role of the water snow line for the formation of rock-rich and ice-rich planetesimals. In this paper, we focus on the formation of rock-rich planetesimals. A companion paper examines the combined formation of both rock-rich and ice-rich planetesimals.  
}
{
We develop a new Monte Carlo code to calculate the radial evolution of silicate particles in a turbulent accretion disk, accounting for the back-reaction (i.e., inertia) of the particles on their radial drift velocity and diffusion. Results depend in particular on the particle injection width (determined from the radial sublimation width of icy pebbles), the pebble scale height and the pebble mass flux through the disk. The scale height evolution of the silicate particles, which is the most important 
factor for the runaway pile-up, is automatically calculated in this Lagrange method.  
}
 {
From the numerical results, we derive semi-analytical relations for
the scale height of the silicate dust particles and the particles-to-gas density ratio at the midplane,
as functions of a pebble-to-gas mass flux ratio and the $\alpha$ parameters for disk gas accretion and vertical/radial diffusion.
We find that the runaway pile-up of the silicate particles (formation of rocky planetesimals) 
occurs if the pebble-to-gas mass flux ratio is 
$\ga [(\Dz/\acc)/3 \times 10^{-2}]^{1/2}$ where $\Dz$ and $\acc$ are the $\alpha$ parameters for vertical turbulent diffusion and disk gas accretion.
 }
{}
     \titlerunning{Pile up of silicate dust particles from sublimating pebbles}
   \maketitle
%

\section{Introduction}

In order for planets to form, micron-size grains must grow to planetesimal size (i.e., kilometers or more) before they are lost by gas drag into the central star. One possibility is through streaming instabilities (SI), a mechanism to concentrate small solid particles to high enough densities to trigger gravitational collapse \citep[e.g.,][]{Youdin2005}.  
A midplane particle-to-gas ratio of order unity is required to trigger strong clumping, 
but this condition is not easy to achieve, even when considering the most favorable particles, i.e. icy pebbles \citep[e.g.,][]{Krijt2016}. It is even more difficult to achieve for smaller silicate dust particles, especially when other sources of turbulence are present \citep{Carrera2015,Yang2017,Gole2020}. 

The role of ice lines, and particularly of the water snow line, has long been recognized as potentially important: The sublimation of pebbles across the snow line leads to an outward diffusion of vapor and its recondensation, 
thus increasing the pebble surface density 
\citep[e.g.][]{Lunine+Stevenson1985, Ciesla+Cuzzi2006, Schoonenberg2017, Drazkowska17}. 
Recondensation of outwardly diffused water vapor 
onto pebbles also induces local pebble growth \citep{Ros2013, Schoonenberg2017, Drazkowska17}. 
A local pressure bump due to the ionization rate change across the snow line
caused by the sublimation of icy pebbles
could halt the pebble drift to result in their pile-up, under favorable disk conditions
\citep{Kretke2007,Brauer2008,IdaLin2008}.
These mechanisms potentially lead to the formation of ice-rich planetesimals\footnote{Our subsequent papers, Hyodo et al. (2020, submitted: referred to as Paper II) and Hyodo et al. (2020, submitted), investigate 
pile-ups of both icy pebble and silicate dust particles.}.
But in the present work, our aim is to show that the sublimation of icy pebbles can also result in 
the formation of silicate-rich (rocky) planetesimals inside the snow line.

Many small silicate dust particles that are released by the sublimation of individual icy pebbles are coupled with the disk gas (Section \ref{subsec:St_sil}).
Because incoming icy pebbles drift fast by gas drag and 
pebbles are supplied with a relatively high flux,
the silicate dust particles
can also pile up inside the snow line, potentially leading to the conditions for a formation of silicate-rich planetesimals by direct gravitational instability \citep{Saito_Sirono2011, Ida_Guillot2016}. Determining the reality of the processes and their relative contributions is of course crucial to understand the formation of planetesimals, the composition of planets and ultimately account for the global ice-to-rock ratio in the solar system \citep[see][]{Kunitomo+2018}. 

However, the studies of this process \citep{Ida_Guillot2016, Schoonenberg2017,Drazkowska17, Hyodo2019} lead to different, seemingly contradictory results. In order to examine the problem in a new light, we develop two tools. A first one is a 2D  ($r$-$z$) Monte Carlo code to simulate the pile-up of small silicate particles injected from sublimating icy pebbles in a turbulent protoplanetary disk, and using the Lagrangian advection-diffusion method by \citet{Ciesla2010,Ciesla2011} with our new addition of the back-reaction to the radial velocity 
and diffusion of the silicate particles. A second one is a 1D diffusion-advection grid simulation based on the work of \citet{Schoonenberg2017} as updated by \citet{Hyodo2019} and including input results obtained from our Monte Carlo code.
These tools enable us to study the fate of silicate dust particles and icy pebbles for a wide range of conditions, including the possibilities that turbulent mixing may differ in the radial and vertical directions \citep{Zhu2015, Yang2018}.
 
We first focus on the 2D Monte Carlo code and its results. In a companion paper (Paper II), we will apply the results to the diffusion-advection simulation, allowing us to include the formation of both silicate-rich and water-rich planetesimals. We will treat the case of complex protoplanetary disk models in a third paper. 

The present article is organized as follows:  In Section 2, we summarize the results of \citet{Ida_Guillot2016}, \citet{Schoonenberg2017}, and
\citet{Hyodo2019}, focusing on the silicate dust particle pile-up near the snow line. 
In Section 3, we describe our Monte Carlo simulation code.
In Section 4, after testing our simulations, we present our results and derive semi-analytical relations for the silicate dust scale height and the pile-up.
We conclude in Section 5.

\section{Previous studies and analytical solution}

\subsection{Setting}

The outskirts of protoplanetary disks are a large reservoir of condensed particules, initially in the form of submicron-sized particles. Their progressive growth and drift due to gas drag \citep{Adachi1976,Weidenschilling1977} can lead to a wave of ``pebbles'', i.e., centimeter- to meter-sized particles that drift rapidly inward from the outer to the inner regions of the disk \citep[e.g.][]{Garaud2007,Lambrechts2014}. The mass flux of incoming pebbles can increase significantly above the standard value of $\sim 1/100$ the gas mass flux that would be obtained for micron-size dust and a solar composition gas. For example, \citet{Appelgren+2020} obtain that, a decrease in the drift rate of pebbles from the outer to the inner region can lead to a $\sim 4$ times increase in the dust-to-gas ratio. Similarly, \citet{Mousis+2019} obtain an enrichment factor of the inner disk in volatiles that can reach $\sim 20$ to $30$. 
\begin{figure}[htb]
\includegraphics[width=95mm,angle=0]{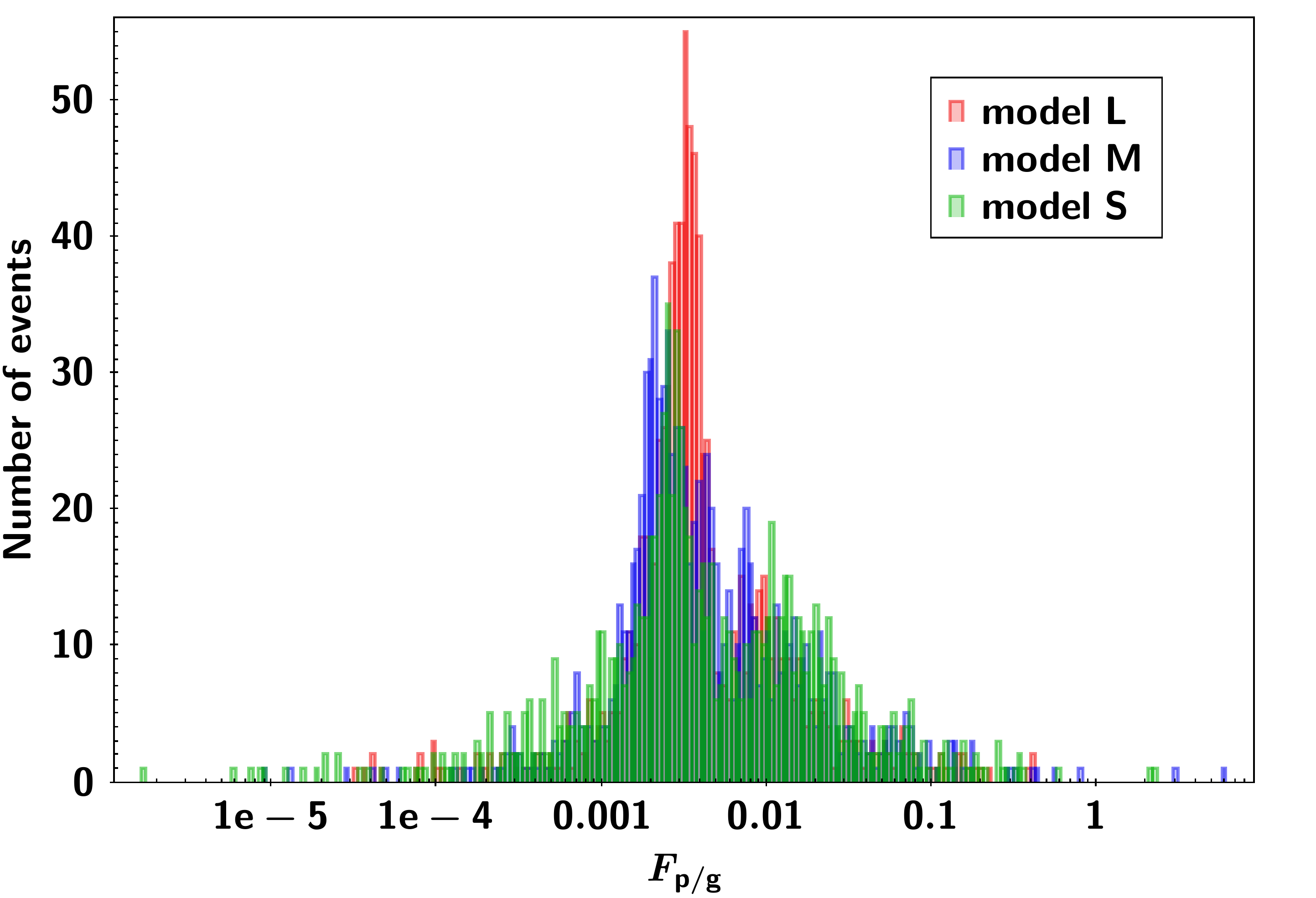}
\caption{  
Histogram of the time-dependent values of $F_{\rm p/g}\equiv\dot{M}_{\rm p}/\dot{M}_{\rm g}$, the ratio of the pebble to gas mass flux at the inner boundary of three protoplanetary disks (models L, M and S, respectively) from \citet{Elbakyan+2020}. For this setting, irregularities in the radial and azimuthal structures of the disks yield a highly time-dependent $F_{\rm p/g}$ value. 
}
\label{fig:elbakyan2020}
\end{figure}
The values of $F_{\rm p/g}$, 
the ratio of the mass flux of pebbles $\dot{M}_{\rm p}$ to that of the gas $\dot{M}_{\rm g}$, depend on the disk models and underlying assumptions on particle growth. 
Using 2D simulations of the formation of a protoplanetary disk from the collapse of a molecular cloud core, \citet{Elbakyan+2020} obtained a highly time-dependent $F_{\rm p/g}$
that is shaped by the perturbations of the disk, as exemplified in  
Fig.~\ref{fig:elbakyan2020}.
The high values, much beyond the canonical value ($\sim 1/100$) are therefore possible
during the time-variation in each disk, either as a progressive increase during the later evolution phases of the disk \citep{Garaud2007, Mousis+2019, Appelgren+2020}, or as short pulses \citep{Elbakyan+2020}, presumably due to the sudden release of pebbles. 

Given the possible presence of such a pebble flux, \citet{Saito_Sirono2011} had pointed out that small silicate dust particles released by incoming icy pebbles would pile up, potentially leading to the conditions for a direct gravitational collapse of the dust subdisk. However, they had assumed a stationary disk without inward disk gas accretion nor diffusion. Our goal is to understand, depending on the value of $F_{\rm p/g}$, whether indeed planetesimals may form from such a pile-up. 

\subsection{A comparison of previous studies}

Several studies have thus far considered the problem in the framework of an evolving accretion disk. 
Using an analytical approach, \citet{Ida_Guillot2016} found that even though the disk gas accretion tends to smooth out 
the radial distribution of silicate dust particles, a runaway pile-up may be achieved because of 
the back-reaction to inward drift of the particles, that is,
the slowing-down of the radial drift of the silicate particles due to the increasing inertia of the
piled-up particles.
They derived a simple analytical criterion linking the critical pebble-to-gas mass flux ratio for runaway pile-up to a ratio between the dust-to-gas scale height and the fraction of dust contained by icy pebbles (Section 2.3 and 4.1.2). 
Because \citet{Ida_Guillot2016} assumed that 
the scale height of silicate particles is similar to that of incoming icy pebbles (i.e., much smaller than the disk gas scale height),
and neglected radial and vertical turbulent diffusion,
their critical pebble accretion flux corresponds to the most favorable limit to pile-up silicate dust particles and form rock-rich planetesimals by gravitational instability. 

\citet{Schoonenberg2017} performed detailed 1D diffusion-advection simulations
of gas and icy pebbles with an Eulerian grid code to study pile-ups of the icy pebbles in an accretion disk, taking account of
the back-reaction to their drift velocity, sublimation process, and 
both radial and vertical turbulent diffusion diffusion.
Near the snow line, re-condensation of water vapor diffused from
the region inside the snow line enhances the ice surface density.  
For icy pebbles with Stokes number of $10^{-2}-1$,
SI can occur for the solid-to-gas surface density ratio $Z_\Sigma \ga 0.02-0.03$
\citep{Johansen2009,Carrera2015,Yang2017}.
They found that
the ice surface density is enhanced by a factor of several by the
recycling process in the case of relatively vigorous turbulence ($\alpha \sim 10^{-3}-10^{-2}$)
and a relatively high pebble-to-gas mass accretion rate ($\fpg \sim 0.8$).

\citet{Schoonenberg2017} also calculated the 
radial distribution of silicate particles released from
drifting pebbles inside the snow line.
However, because they were focusing on the pebbles' SI,
they did not consider the back-reaction to
the radial drift velocity of the silicate particles.
They also assumed immediate turbulent stirring
such that the scale height of silicate particles is 
the same as the disk gas scale height.
As a result, they did not find a runaway pile-up of the silicate particles
(for details, see Section 2.3 and 4.1.2).
 
\citet{Hyodo2019} updated 
\citet{Schoonenberg2017}'s 1D diffusion-advection grid simulations,
adding the back-reactions to radial drift and diffusion of silicate particles.
They also considered the evolution of the scale height of silicate dust particles.
When the silicate particles are released from icy pebbles passing the snow line,
their scale height is similar to that of icy pebbles.
It is gradually increased by the vertical turbulent diffusion.
\citet{Hyodo2019} analytically modeled the dust scale height evolution and incorporated the model
into the grid simulation.
They treated the diffusion $\alpha$-parameter and 
the effective $\alpha$-parameter for the disk accretion independently.
The turbulent mixing and the angular momentum transfer
can be different mechanisms. 
For example, the angular momentum transfer of the disk can be 
dominated by a process other than radial turbulent diffusion such as disk winds \citep[e.g.,][]{Suzuki16,Bai16}.
Therefore, it is reasonable to distinguish the diffusion and the accretion $\alpha$-parameters. 
They identified the parameter regions of the runaway pile-up of the silicate particles
on the plane of the pebble mass flux and the diffusion $\alpha$-parameter.

As \citet{Ida_Guillot2016} suggested, the silicate particle pile-up condition
depends directly on their scale height.
\citet{Hyodo2019} estimated the scale height evolution only by accounting for vertical diffusion, and not taking radial diffusion into account.

\subsection{Analytical derivation}

As we described in Section 1,
we assume that icy pebbles consists of 
a large number of small silicate particles covered by icy mantles.
When the pebbles drift inward and pass the snow line,
the icy mantles are sublimated and consequently the small silicate particles are released.  
We set that the Stokes number of icy pebbles is $\stpeb = 0.1$, which is relevant to the Epstein regime
\citep{Okuzumi2012,IGM16}, and that of silicate particles is $\st \ll 1$,
where the Stokes number is defined by the stopping time due to gas drag $t_{\rm stop}$
and Keplerian frequency $\Omega_{\rm K}$ as $\tau_{\rm s} = t_{\rm stop} \Omega_{\rm K}$.

We consider a steady accretion disk.
We show that our formulation and the results are scaled by the pebble-to-gas mass flux ratio
and we do not need to specify the magnitude of the disk gas accretion rate.
Actually, the pebble mass flux through the disk ($\mdotp$) is proportional to disk gas accretion rate ($\mdotg$)
in the case of the steady accretion before
the pebble formation front reaches the disk outer edge \citep{IGM16}. 
\citet{Kanagawa2017} pointed out that the vertically averaged disk gas flow
can be outward,
when $\mdotp$ is so large that the advective angular momentum carried by pebbles
dominates over the viscous angular momentum transfer.
In that case, the steady accretion disk model breaks down.
However, as we show in Appendix A,
such a situation is not realized in the parameter range we cover in this paper. 

\begin{table}
\caption{List of notations}
\begin{tabular}{c|cc}  \hline
notation    & definition & eq. \# \\ \hline \hline
$f_{\rm d/p}$ & silicate dust mass fraction in a pebble & \\ 
                           & beyond the snow line [$= 0.5$ (nominal)]&     \\ \hline
$F_{\rm p/g}$ & pebble-to-gas mass flux in the disk &  (\ref{eq:Fpg}) \\
                            & $(= \mdotp/\mdotg)$  & \\ \hline 
$Z_\Sigma$     & vertically averaged dust-to-gas mass ratio & (\ref{eq:Z_Sigma})  \\
                           & $(= \sigd/\sigg)$ & \\ \hline
$Z$                    & local dust-to-gas mass ratio  & (\ref{eq:Z})  \\
                           &  ($= \rhod/\rhog = Z_\Sigma/\hdg$) &     \\   \hline
$\Lambda$      & parameter for the solid particle inertia &   (\ref{eq:Lambda}) \\
                          & $[=\rhog/(\rhod+\rhog) = 1/(1+Z)]$ & \\ \hline
$C_\eta$          & power index of pressure gradient & (\ref{eq:C_eta}) \\
                           & $[= -(1/2)(\partial \ln P/\partial \ln r); = 11/8$(nominal)] & \\ \hline
$\acc$              & $\alpha$-parameter for disk gas accretion & (\ref{eq:u_nu}) \\ 
$\Dr$                & $\alpha$-parameter for radial mixing &  \\ 
$\Dz$                & $\alpha$-parameter for vertical stirring & \\  \hline
$(\rhod/\rhog)_0$ & maximum value of $\rhod/\rhog$ in all the grids & \\
                           &  (usually, at the grid of $(x,z) \sim (0,0)$) & \\ \hline
$\hdg$              & dust-to-gas scale height ratio $(=\hd/\hg)$ & (\ref{eq:H_d}) \\ 
$h_{\rm d/g,0}$  & $\hdg$ at $x$ with $(\rhod/\rhog)_0$ \\ 
$\hpg$              & pebble-to-gas scale height ratio $(=\hp/\hg)$ & (\ref{eq:H_peb}) \\ 
\hline
$\stpeb$           & Stokes number of pebbles [$=0.1$ (nominal)] \\
$\st$                 & Stokes number of dust [$=10^{-5}$ (nominal)] \\
\hline
\end{tabular}
\end{table}

First we derive the vertically averaged solid-to-gas mass ratio (metallicity),
inside the snow line. 
In this paper, the subscripts, ``g",``p", and ``d" represent the disk H-He gas,
icy pebbles, and silicate dust particles, respectively.
In the steady state, the surface density of the silicate dust particles $\sigd$
and the disk gas $\sigg$ are given by 
\begin{align}
\sigd & = f_{\rm d/p} \mdotp/2\pi r \varv_r,  \\
\sigg & = \dot{M}_{\rm g}/2\pi r u_r,  
\label{eq:Sigmag}
\end{align}
where $\vr$ and $u_r$ are the drift velocity
of the silicate particles and the gas accretion velocity,
and $f_{\rm d/p}$ is the silicate dust mass fraction in the icy pebbles.
We adopt $f_{\rm d/p}=0.5$ as a nominal value.
The vertically averaged metallicity is
\begin{equation}
Z_\Sigma \equiv \frac{\sigd}{\sigg} = f_{\rm d/p} F_{\rm p/g} \frac{u_r}{\vr},
\label{eq:Z_Sigma}
\end{equation}
where 
\begin{equation}
F_{\rm p/g} \equiv \frac{\mdotp}{\dot{M}_{\rm g}}.
\label{eq:Fpg}
\end{equation}
We also use the local solid-to-gas ratio defined by
\begin{equation}
Z \equiv \frac{\rhod}{\rhog},
\label{eq:Z}
\end{equation}
where $\rhog$ and $\rhod$ are the local densities of gas and silicate dust particles.

The radial drift velocity of the particles and disk gas is given by
\citet{Ida_Guillot2016} and \citet{Schoonenberg2017},
\begin{align}
\vr & = -\Lambda^2 \frac{2\st}{1+\Lambda^2 \st^2}\eta \vk
- \Lambda \frac{1}{1+\Lambda^2 \st^2}u_\nu, \label{eq:vr} 
\end{align}
where $\vk$ is Keplerian velocity,
$\st$ is Stokes number of the particles,
$\eta$ is the degree of deviation of the gas rotation angular velocity 
($\Omega$) from
Keplerian one ($\Omega_{\rm K}$), given by
\begin{align}
\eta & \equiv \frac{\Omega_{\rm K} -\Omega}{\Omega_{\rm K}}
= C_\eta \left(\frac{\hg}{r}\right)^2, \label{eq:eta}\\
C_\eta & \equiv \frac{1}{2} \frac{\partial \ln P}{\partial \ln r}, \label{eq:C_eta}
\end{align}
where $C_\eta = 1.3-1.4$, depending on the disk structure (Ida et al. 2016).
We use $C_\eta = 11/8$ for $\sigg \propto r^{-1}$ and $T \propto r^{-1/2}$
in this paper. 
In the above equation, $u_\nu$ is an unperturbed disk gas accretion (advection) velocity given by
\begin{eqnarray}
u_\nu \simeq \frac{3\nu_{\rm acc}}{2r} 
         \simeq \frac{3 \acc H_{\rm g}^2 \Omega}{2r} 
          \simeq \frac{3\acc}{2} \left(\frac{H_{\rm g}}{r}\right)^2 \vk, \label{eq:u_nu}
\end{eqnarray}
where $\nu_{\rm acc}$ is the effective viscosity for the angular momentum transfer of the disk gas,
$\hg$ is the disk gas scale height, and $u_\nu$ is defined to be positive for inward flow.
The effect of back-reaction from the piled-up particles
is represented by $\Lambda$ that is 
defined by
\begin{equation}
\Lambda \equiv \frac{\rhog}{\rhog + \rhod}=\frac{1}{1+Z}.
\label{eq:Lambda}
\end{equation}
Because $\rho_{\rm d}\simeq \Sigma_{\rm d}/\sqrt{2\pi} H_{\rm d}$,
where $\hd$ is the scale hight of the silicate dust particles,
and $\rho_{\rm g}\simeq \Sigma_{\rm g}/\sqrt{2\pi} H_{\rm g}$,
Eq.~(\ref{eq:Lambda}) for $z < H_{\rm d}$ is approximated for the particle motion by
\begin{equation}
\Lambda^{-1} = 1 +  \frac{\rho_{\rm d}}{\rho_{\rm g}}
\simeq 1 + Z_\Sigma \frac{H_{\rm g}}{H_{\rm d}}
\simeq 1 + h_{\rm d/g}^{-1} Z_\Sigma.
\label{eq:Lambda2}
\end{equation}
We define the scale height ratio, 
\begin{equation}
h_{\rm d/g} \equiv \hd/\hg.
\label{eq:H_d}
\end{equation}

In the case of silicate particle pile-up due to sublimation of icy pebbles,
the pile-up is radially localized just inside of the snow line.
If we apply Eq.~(\ref{eq:ur2}), the gas accretion velocity ($u_r$) is locally slowed down from $u_\nu$
by the back-reaction to the gas motion.
If the steady accretion with constant $\acc$ is assumed,
the gas surface density given by Eq.~(\ref{eq:Sigmag}) is locally increased.
However, in reality, the assumption for vertically constant $\acc$ may be broken down
in that case,
or some local instability may smooth out the local gas concentration.
Therefore, we assume that $u_r \simeq u_\nu$ also in the silicate dust pile-up region.
Even if we take account of the back-reaction to the gas motion, the condition for 
the runaway pile-up of the silicate dust particles does not change significantly, as shown 
in Appendix A. 

Because $\st \ll 1$, silicate particle motions are strongly coupled with disk gas,
but the back-reaction to the drift velocity of the silicate particles must be be taken into account.
Equation~(\ref{eq:vr}) with $\st \ll 1$ implies $\vr \simeq \Lambda u_\nu$.
Substituting $u_r \simeq u_\nu$, $\vr \simeq \Lambda u_\nu$,
and Eq.~(\ref{eq:Lambda2}) into Eq.~(\ref{eq:Z_Sigma}),
we obtain
\begin{eqnarray}
Z_\Sigma \simeq \frac{f_{\rm d/p}}{\Lambda} \, F_{\rm p/g} 
      \simeq f_{\rm d/p}(1+\hdg^{-1} Z_\Sigma) F_{\rm p/g},
\label{eq:Zs}
\end{eqnarray}
which is solved as \citep{Ida_Guillot2016}
\begin{eqnarray}
Z_\Sigma =\frac{f_{\rm d/p} F_{\rm p/g}}{1- \hdg^{-1} f_{\rm d/p} F_{\rm p/g}}.
\label{eq:IG2}
\end{eqnarray}
The midplane $Z=\rho_{\rm d}/\rho_{\rm g}$ is given by
\begin{eqnarray}
Z = \frac{Z_\Sigma}{h_{\rm d/g}} =
\frac{f_{\rm d/p} F_{\rm p/g}}{h_{\rm d/g}  - f_{\rm d/p} F_{\rm p/g}}.
\label{eq:IG3}
\end{eqnarray}
If we consider the back-reaction to gas motion,
the divergence condition is only slightly modified to (see Appendix A):
\begin{align}
Z = \frac{f_{\rm d/p} F_{\rm p/g}}{h_{\rm d/g}  - (1-\hpg) f_{\rm d/p} F_{\rm p/g}}.
\label{eq:IG4}
\end{align}
The metallicity $Z$ diverges, 
which would lead to rocky planetesimal formation, for
 \begin{align}
F_{\rm p/g} & > \frac{h_{\rm d/g}}{ f_{\rm d/p}}, 
\hspace{0.5cm} [\mbox{w/o the back rection to gas}] \\
F_{\rm p/g} & >  \frac{h_{\rm d/g}}{(1-\hpg) f_{\rm d/p}}. \hspace{0.5cm} [\mbox{w/ the back rection to gas}]
\label{eq:IG4}
\end{align}
The former condition was derived by \citet{Ida_Guillot2016}.
We mostly refer to the former condition.  

\citet{Schoonenberg2017} neglected the back-reaction to the silicate dust 
drift velocity and assumed $\Lambda = 1$ in Eq.~(\ref{eq:Zs}).
In this case, Eq.~(\ref{eq:Zs}) is reduced to $Z_\Sigma \simeq f_{\rm d/p} F_{\rm p/g}$,
which shows no divergence.
This demonstrates that the back-reaction 
to the silicate particle radial velocity plays an essential role
in the occurrence of the runaway pile-up.

In the runaway pile-up condition, $F_{\rm p/g} > f_{\rm d/p}^{-1} \,\hdg$, 
the estimation of $\hd$ is essentially important.
The scale height of pebbles ($\hp$) is given with their Stokes number $\tau_{\rm s,p}$ and
the vertical mixing parameter $\Dz$ as \citep{Dubrulle1995,Youdin2007}
\begin{eqnarray}
\hpg =  \frac{\hp}{\hg} = \left(1+ \frac{\tau_{\rm s,p}}{\Dz}\right)^{-1/2}.
\label{eq:H_peb}
\end{eqnarray}
We will discuss the increase of $\hp$ by Kelvin-Helmholtz instability 
for the vertical shear due to pile-up of particles in Paper II.

The scale height of silicate dust particles ($H_{\rm d}$) must be 
the same as $H_{\rm p}$ just after the release from the icy pebbles.
Because $\st \ll \tau_{\rm s,p}$, $H_{\rm d}$ is increased by the vertical turbulent stirring afterward.
\citet{Ida_Guillot2016} assumed that $H_{\rm d} \simeq \hp$
($h_{\rm d/g} = h_{\rm p/g}$), until the runaway pile-up develops.
For $h_{\rm d/g} \sim 0.1$ ($\tau_{\rm s,peb}/\Dz \sim 10^{2}$) and $f_{\rm d/p}  \sim 0.5$, 
the optimistic estimate by Eq.~(\ref{eq:IG3}) 
predicts that the runaway pile-up
occurs for $F_{\rm p/g} > 0.2$, which would be available for relatively early phase of disk evolution. 

On the other hand, \citet{Schoonenberg2017}
assumed immediate vertical stirring by turbulence, that is, $H_{\rm d} \simeq H_{\rm g}$
($h_{\rm d/g} \simeq 1$).
Because they neglected the back-reaction to $\vr$,  they ever found the runway pile-up in their simulations.
Even if we apply 
 Eq.~(\ref{eq:IG3}), which was derived with the back-reaction, for the case of $h_{\rm d/g} \simeq 1$
 and $f_{\rm d/p}  \sim 0.5$, the runaway pile-up condition is $F_{\rm p/g} > 2$. 
 The high value of $\fpg$  may not be completely ruled out, but not easy to be established \citep[][and see Fig.~\ref{fig:elbakyan2020} in this paper]{Ida_Guillot2016}.

\citet{Hyodo2019} treated the effects of the vertical and radial diffusion more carefully.
They considered the vertical stirring characterized by $\Dz$ to derive a model for 
$h_{\rm d/g}$ as a function of $x=r-r_{\rm snow}$.
They also introduced 
the back-reaction to the diffusion coefficients $D_r$ and $D_z$ by the pile-up of the particles such that
\begin{align}
D_r & = \Dr \hg^2\Omega \Lambda^K, \label{eq:KDr}\\
D_z & = \Dz \hg^2\Omega \Lambda^K. \label{eq:KDz}
\end{align}
A simple argument that $D_z, D_r \propto \hg^2 \propto c_s^2
\propto P/(\rhog + \rhod) \propto \Lambda$ 
\citep{Laibe2014,LinYoudin2017} suggests $K=1$.
Another argument is based on Kolmogorov theory
\citep[e.g.][]{Cho2003}.
In the eddy cascade,  the energy transfer rate, $(\rhog + \rhod) \, \varv_\ell^2 /t_{\rm eddy}$ is 
independent of $\ell$, 
 where $\ell, \varv_\ell$, and $t_{\rm eddy}$ are the eddy size,
 velocity, and turnover time $\sim \ell/\varv_\ell$.
Because $D_z, D_r \sim \ell_0 \varv_{\ell 0}$ where $\ell_0$ and $\varv_{\ell0}$
are the eddy size and velocity controlling the turbulence,
$D_z, D_r \propto \ell_0 \,[\ell_0/(\rhog + \rhod)]^{1/3} \propto \Lambda^{1/3}$. 
This argument thus implies $K=1/3$. This should be a lower limit because an increase 
in the solid-to-gas is expected to suppress turbulence itself. 
Given these uncertainties,
\citet{Hyodo2019} performed runs with $K=$ 0, 1, and 2.
They found that the runaway pile-up is caused by inclusion of
the back-reaction to $\vr$, and that the condition for the runaway pile-up
depends on the relations among $\acc$, $\Dr$ and $\Dz$, while
\citet{Ida_Guillot2016} considered the limits of $\Dr, \Dz \rightarrow 0$ 
and their condition, Eq.~(\ref{eq:IG3}), does not include $\acc$.
The condition derived by \citet{Hyodo2019}
is more severe than the prediction by \citet{Ida_Guillot2016}.
The runaway pile-up
occurs for $F_{\rm p/g} > 0.2-0.3$ and $\Dr = \Dz \simeq 10^{-3}$
in the case of $K= 1$ and 2 with $\acc=10^{-2}$, 
while it does not occur with $K=0$ as long as
in the parameter space of $\Dr = \Dz = 10^{-3}-10^{-2}$ and $F_{\rm p/g} < 0.6$.
In this paper, we will show this back-reaction effect on the diffusion coefficients more clearly. 

\subsection{Stokes number of silicate particles} 
\label{subsec:St_sil}

The models discussed thus far and the model that we present are based on the assumption that 
many silicate dust particles released by the individual icy pebbles remain small compared to pebbles (i.e., with a Stokes number that is much smaller than unity). 
In this case, the radial drift is much faster for icy pebbles than for silicate particles.
Because the gas accretion velocity is given by $u_\nu \sim (3/2) \acc (\hg/r)^2 \vk$
(Eq.~(\ref{eq:u_nu})) and the drift speed relative to the gas
is given by $\varv_r \simeq \st (\hg/r)^2 \vk$ (Eq.~(\ref{eq:vr})),
the total drift velocity is dominated by $u_\nu$ if $\st \la \acc$.
Therefore, the assumption here is expressed as $\st \la \acc < \stpeb$.

In Section 2.3, we assumed that $\stpeb \sim 0.1$, $\acc \sim 10^{-3}-10^{-2}$, and $\st \la \acc$. 
\citet{Saito_Sirono2011} assumed that many $\mu$m-sized silicate particles are embedded in pebbles.  
\citet{Morbidelli15} considered chondrule-sized ($\sim$ mm) particles.  
Because the Stokes number is proportional to the particle size in the Epstein regime
and its square in Stokes regime, it is reasonable to assume that $\st < \acc$.
This model corresponds to the ``many-seeds model" of \citet{Schoonenberg2017}.
In the Monte Carlo simulations here, we adopt $\st = 10^{-5}$.

In the silicate particle pile-up region,
the particles may quickly grow up to sizes determined by the threshold collision velocity for fragmentation/rebound. 
The collision velocity is set by 
the velocity dispersion induced by the turbulence $\sim (3 (\Dr^2+\Dz^2)^{1/2} \st)^{1/2} c_s$
or by the drift velocity difference between the particles $\sim 2\st \, \eta \, \vk$ \citep[e.g.,][]{Sato2016}.
The collision velocity between silicate particles
is therefore $(\st/\tau_{\rm s,p})^{-(0.5-1)}$ times of that between icy pebbles.
If the threshold collision velocity is 10 times lower for the silicate particles 
than for the icy particles \citep{Blum00, Zsom+2010,Zsom11,Wada11,Weidling12,Wada13},
the silicate particles can grow only up to sizes such that
$\st \sim (10^{-1}$--$10^{-2})\tau_{\rm s,p}
\sim 10^{-3}$--$10^{-2}$.
Therefore, for $\acc = 10^{-3}$--$10^{-2}$,
the drift velocity of the silicate particles is comparable to $u_\nu$ and
the pile-up would not be significantly affected by the value of $\st$.

We note that this conventional view has been challenged recently \citep{Kimura2015,Gundlach2018,Musiolik2019,Steinpilz2019}. A higher fragmentation threshold for silicates would open the possibility for dust to coagulate and grow. 
We hence performed runs with larger values of $\st$
to find that the results are not affected for $\st \la \acc$
and that the runaway pile-up region is removed from
lower $\fpg$ regions as $\st$ becomes larger.

Our model hence remains valid if the fragmentation velocity for rock-rich dust remains smaller than for ice. We also note that an exploration of the consequence of a different change of fragmentation velocity across ice lines is explored by  \citet{Vericel+2019} \citep[see also][]{Gonzalez+2017}. In this case, planetesimal formation may still be possible, but through a slow-down of the drift rate due to the gas back-reaction and to a progressive growth of the particles.

\section{The Monte Carlo approach}

In order to model precisely the effects of drift due to gas drag, gas advection, radial and vertical diffusion, we develop a Monte Carlo simulation of silicate dust particles in a 
turbulent accretion disk. The dust particles released from sublimating icy pebbles are injected near the snow line.
The back-reactions to $\vr$ and the diffusion of the particles are included using the super-particle approximation.
We set radial and vertical coordinates, $(x,z)$, where $x \equiv r - r_{\rm snow}$
is the radial distance from the snow line at $ r_{\rm snow}$ 
and $z$ is the distance from the midplane.

We consider a radially local region near the snow line
and neglect the $r$-dependence of the Keplerian frequency ($\Omega$) and the disk gas scale height ($\hg$). 
As discussed in Section 2.4, we adopt the conventional view to set $\stpeb = 0.1$ and $\st = 10^{-5}$ in our simulations.

\subsection{Injection of particles} 

To model the release from drifting icy pebbles due to sublimation,
at each time step $\delta t$, we randomly inject a new silicate super-particle with 
mass $m$ with a radially uniform distribution near the snow line
in the range of $- 0.5 \Delta x_{\rm subl}  < x  < 0.5 \Delta x_{\rm subl}$ and 
a Gaussian distribution of $z$ of the root mean square $\Delta z_{\rm subl}$ as
\begin{align}
x & = 0.5 \;{\cal R}\; \Delta x_{\rm subl}, \\
z & = \sqrt{2} \; {\rm erf}^{-1}(|{\cal R}|) \;\frac{{\cal R}}{|{\cal R}|}\Delta z_{\rm subl},
\end{align}
where ${\cal R}$ is a random number in the range of $[-1,1]$,
$\Delta x_{\rm subl}$ is the characteristic sublimation radial length,
and $\Delta z_{\rm subl}$ corresponds to 
the scale height of the incoming icy pebbles, $\hp$ given by Eq.~(\ref{eq:H_peb}).
We adopt $\Delta x_{\rm subl}=0.1\hg$ as a nominal case.
In Section 4.4.2, we also take a more complicated function
that fits the numerical result of the grid code simulation obtained by Paper II.

Here we distinguish the effective viscosity parameters for advection, radial mixing, and
vertical mixing, denoted by $\acc$, $\Dr$, and $\Dz$, respectively.
According to different values of $\Dz$, we use the consistent value of 
the scale height of the injected silicate particles given by Eq.~(\ref{eq:H_peb}).
Although \citet{Hasegawa2017} suggested $\Dr, \Dz \sim 0.1 \, \acc$,
the relations among $\Dr, \Dz$ and $\acc$ are not clear \citep[e.g.,][]{Armitage2013}.
Therefore, we survey broad parameter ranges of $\Dr, \Dz$ and $\acc$; 
We mostly show the results with $\acc \ge \Dr = \Dz$.  
In some cases, the results with $\Dr \neq \Dz$ ($\acc \ge \Dr, \Dz$) are also shown.

\subsection{Advection and diffusion} 

At each time step $\delta t$, we change the $r$--$z$ locations of  
the particles that were injected before, following \citet{Ciesla2010,Ciesla2011} by 
\begin{align}
\delta x & = v_{{\rm adv},r} \delta t  + {\cal R} \,(6 D_r \delta t)^{1/2}, \label{eq:dx}\\ 
\delta z & = v_{{\rm adv},z} \delta t  + {\cal R} \, (6 D_z \delta t)^{1/2}, \label{eq:dz}
\end{align}
where the 1st and 2nd terms in the right hand side represent 
advection (drift) and diffusion, respectively.
The root mean square of ${\cal R}$ is $1/\sqrt{3}$.
 In our simulations, we adopt $\delta t = \Omega^{-1}$.
\citet{Fromang2006} suggested a turbulent correlation time in
a protoplaneatry disk is $\sim 0.15 T_{\rm K} \sim \Omega^{-1}$.
\citet{Ciesla2011} adopted planar ($x$--$y$) random walks, in addition to
vertical one ($z$), to take account of the global curvature effect.
In our case, because we only consider a local radial range,
we adopt a simpler $r$--$z$ random walks. 

We perform runs with $K=0$ and $K=1$ for the diffusion coefficients
(Eqs.~(\ref{eq:KDr}) and (\ref{eq:KDz})), with $K=1$ being our nominal case.
We also performed runs with $K=1/3$ (the lower limit) and found that the overall features of the results do not depend on 
the value of $K$ (only the runaway pile-up timescale is different),
as long as back-reaction to diffusion is considered (i.e., $K>0$).
Hereafter, we show the results with $K=1$ as representative of the cases including back-reaction to diffusion.

The radial drift (advection) velocity $v_{{\rm adv},r}$ is the same as $\vr$ in Eq.~(\ref{eq:vr}). 
The vertical advection velocity is \citep{Ciesla2010}
\begin{equation}
v_{{\rm adv},z} = - (\Dz + \st)\Omega \; z.
\label{eq:advz}
\end{equation}
The 1st term ($\propto \Dz$) comes from
the effect that diffusion acts to smooth the concentration ($\rhod/\rhog$),
but not $\rhod$ itself.

\subsection{Inclusion of back-reaction effects} 

The advection and diffusion scheme was originally developed by \citet{Ciesla2011} and \citet{Ciesla2011}. Here, we also include the effects of back-reaction
due to the silicate particle pile-up as well as a scheme to model the injection of silicate particles from
sublimating icy pebbles. 

We utilize the ``super-particle" method, 
where one super-particle represents the mass of a large number of particles
while it suffers the same specific drag force as the individual particles.
We inject one super-particle at every timestep $\delta t = \Omega^{-1}$.
For given $\mdotp$, we can determine the individual mass of the super-particles as
\begin{equation}
m = f_{\rm d/p} \mdotp \, \delta t = f_{\rm d/p} \mdotp \, \Omega^{-1},
\end{equation}
where $f_{\rm d/p}$ is a silicate fraction in migrating pebbles
($f_{\rm d/p}=0.5$ in the nominal case).
To locally average the silicate dust mass density, 
we use linear grids, $\Delta x$ and $\Delta z$.
In the nominal cases, we adopt $\Delta x = 0.1\, H_{\rm g}$.
We use $\Delta z \sim \Delta z_{\rm subl}$ near the midplane
and larger $\Delta z$ for upper regions to keep statistically enough 
number of particles in the individual grids.

The silicate particle mass density at a grid of $[x, x+\Delta x]$ and
$[|z|, |z|+\Delta z]$ is 
\begin{equation}
\rho_{\rm d} = \frac{m\, \Delta N_{x,z}}{2\pi r \Delta x \times 2 \Delta z}
= \frac{H_{\rm g}}{r} \frac{\Delta N_{x,z} \, f_{\rm d/p} \mdotp }{4\pi \, (\Delta x/\hg) \, (\Delta z/\hg) \, \hg^3\, \Omega},
\end{equation}
where $\Delta N_{x,z}$ is a total number of the particles in the grid.
We simply assume the vertical isothermal hydrodynamical equilibrium.
Then the gas density is given by
\begin{align}
\rho_{\rm g} & = \frac{\sigg}{\sqrt{2\pi} H_{\rm g}} \exp\left(-\frac{z^2}{2H_{\rm g}^2}\right)
\nonumber \\
 & =\frac{\mdotg}{\sqrt{2\pi} \, 3\pi \alpha_{\rm acc} H_{\rm g}^3 \, \Omega} 
\exp\left(-\frac{z^2}{2H_{\rm g}^2}\right).
\end{align}
The silicate dust to gas ratio at $(x,z)$ is
\begin{align}
Z & = \frac{\rho_{\rm d}}{\rho_{\rm g}} =
\frac{3\sqrt{2\pi}}{4}
\frac{H_{\rm g}}{r} \,\alpha_{\rm acc}
\frac{f_{\rm d/p} \mdotp}{\mdotg} 
\frac{\Delta N_{x,z}}{(\Delta  x/\hg)(\Delta z/\hg)}
\exp\left(\frac{z^2}{2H_{\rm g}^2}\right) \nonumber \\
 & = 3.75 \times 10^{-2} 
\left( \frac{H_{\rm g}/r}{0.04} \right) 
\left( \frac{ \alpha_{\rm acc}}{10^{-2}}\right)
\left( \frac{f_{\rm d/p}}{0.5} \right)
 F_{\rm p/g} \nonumber \\
& \hspace{0.5cm} \times 
\left(\frac{\Delta x}{0.1\hg}\right)^{-1}
\left(\frac{\Delta z}{0.1\hg}\right)^{-1} 
\Delta N_{x,z} \exp\left(\frac{z^2}{2\hg^2}\right).
\label{eq:backreactionZ}
\end{align}
We count $\Delta N_{x,z}$ in the simulation bins
of $[x, x+\Delta x]$ and $[| z |, | z | +\Delta z]$ 
to calculate $Z$ according to Eq.~(\ref{eq:backreactionZ}).
With calculated $Z$, we update $\Lambda = 1/(1+Z)$
in Eqs.~(\ref{eq:dx}) and (\ref{eq:dz}) 
through Eqs.~(\ref{eq:vr}), (\ref{eq:KDr}), and (\ref{eq:KDz})
for the next step.
  
In the simulations, we scale all the length by $\hg$,
including the range of particle injection, $\Delta x_{\rm subl}$ and $\Delta z_{\rm subl}$,
and the grid sizes, $\Delta x$ and $\Delta z$.
The time is scaled by $\Omega^{-1}$, that is, each super-particle
is injected at every unit scaled time.
Therefore, our results can be applied for any location of $r_{\rm snow}$,
as long as $H_{\rm g}/r$ is the same.

In the simulations here, we adopt $f_{\rm d/p}=0.5$ and $\hg/r =0.04$ at $r=r_{\rm snow}$, as nominal values. 
We do not need to specify the values of $\hg$ and $\mdotg$
as well as the value of $r_{\rm snow}$, 
because the only scaled values, $\hdg (=\hd/\hg), \hpg (=\hp/\hg), \hg/r$, and 
$\fpg(=\mdotp/\mdotg)$, are used in the simulations.
For the viscous $\alpha$-parameters,
we will show that only the ratios, $\Dz/\acc$ and $\Dr/\acc$, are important.

\section{Results}

\subsection{Code check}

We test our code by comparing the simulated particle scale heights
with the existing analytical argument, as \citet{Ciesla2010} already did,
and by reproducing the simple analytical result by \citet{Ida_Guillot2016}
and the numerical result without the back-reaction to
silicate particles by \citet{Schoonenberg2017}.

We will also do detailed comparison with the results 
obtained by an updated code of \citet{Hyodo2019} in Paper II.
While \citet{Hyodo2019} included both non-zero $\Dr$ and $\Dz$,
they assumed the evolution of $\hd$ analytically estimated from only the vertical stirring with $\Dz$.
In our simulation, the evolution of $\hd$ is self-consistently calculated
and we found that the radial mixing with $\Dr$ is also important for $\hd$.
On the other hand, the sublimation width $\Delta x_{\rm subl}$ is self-consistently calculated
in \citet{Hyodo2019}, while it needs to be assumed in the simulation here.

\subsubsection{Confirmation of the theoretically predicted particle scale height}

\begin{figure}[htb]
\includegraphics[width=75mm,angle=0]{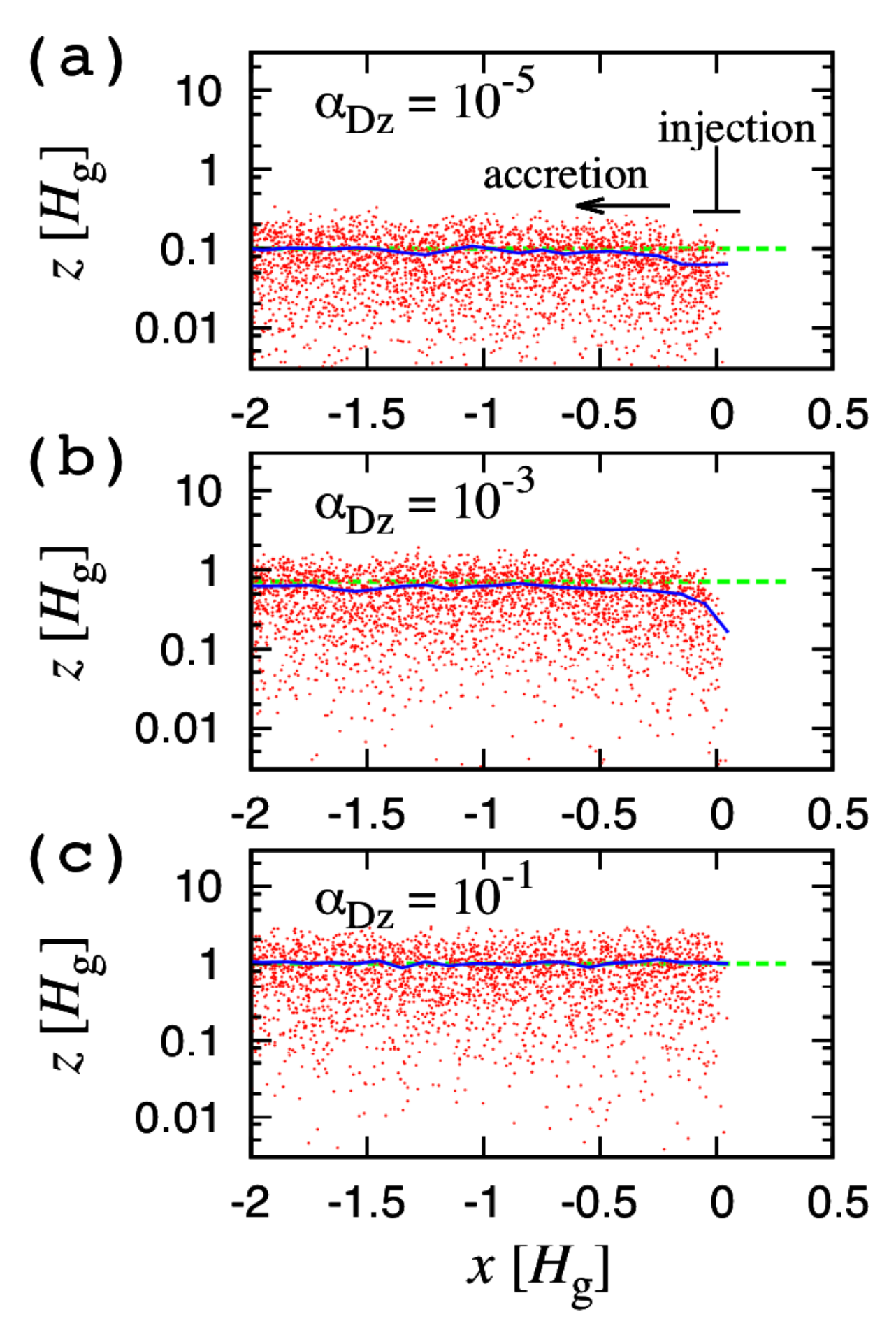}
\caption{  
Snapshots of the particle distribution (red dots) on the $x$-$z$ plane at time $t=10^4 \Omega^{-1}$ 
and the root mean square of $z$ (blue solid curve) as a function of the radial direction $x$ (centered on the snow line location and in units of the gas pressure scale height $H_{\rm g}$). 
We plot $\mid z \mid$, while $z$ takes values of either positive or negative.
We set $\acc = 10^{-2}$, $\Dr=0$ and $\tau_{\rm s} = 10^{-3}$. 
The vertical mixing parameter is (a) $\Dz = 10^{-5}$, (b) $\Dz = 10^{-3}$, and (c) $\Dz = 10^{-1}$. 
The analytical estimates of the equilibrium particle scale height for each $\Dz$
are shown by the dashed green lines.
The back-reactions are not included to $\vr$ ($\Lambda = 1$) and $\Dz$ ($K = 0$), 
We inject the particles in the range of $x = [-0.05, 0.05] \hg$ and $z =  [-0.1, 0.1] \hg$.
}
\label{fig:hd3}
\end{figure}

Figure~\ref{fig:hd3} shows the distribution of particles 
and their scale height
(the local root mean square of $z$ of the particles) in steady state
at $t=10^4 \Omega^{-1}$ (after $10^4$ particles are injected), 
obtained by our Monte Carlo simulations
for (a)  $\Dz = 10^{-5}$, (b) $10^{-3}$, and (c) $10^{-1}$. 
For this runs, we neglected the back-reactions to $\vr$ ($\Lambda = 1$) 
and $\Dz$ ($K = 0$).
To highlight the effect of vertical stirring, we set $\Dr=0$.
The other parameter are the particles' Stokes number $\tau_{\rm s}=10^{-3}$ 
and $\acc = 10^{-2}$.  
The the equilibrium scale height predicted from Eq.~(\ref{eq:H_peb}) 
with $\tau_{\rm s,p}$ replaced by $\tau_{\rm s} = 10^{-3}$
is (a) $0.1 \hg$, (b) $0.71 \hg$ and (c) $1.0 \hg$, respectively,
which are shown in the green dashed lines in the plots.
As the particles drift inward, the local root mean square of $z$
(represented by the blue solid curves) 
asymptotically approaches the theoretical values  in each panel.    
Thus, we confirm that the analytically derived Eq.~(\ref{eq:H_peb}) 
is reproduced by our Monte Carlo simulations.

\subsubsection{Reproduction of \citet{Schoonenberg2017}'s and \citet{Ida_Guillot2016}'s results}
\label{subsec:IG}

\begin{figure}[htb]
\includegraphics[width=90mm,angle=0]{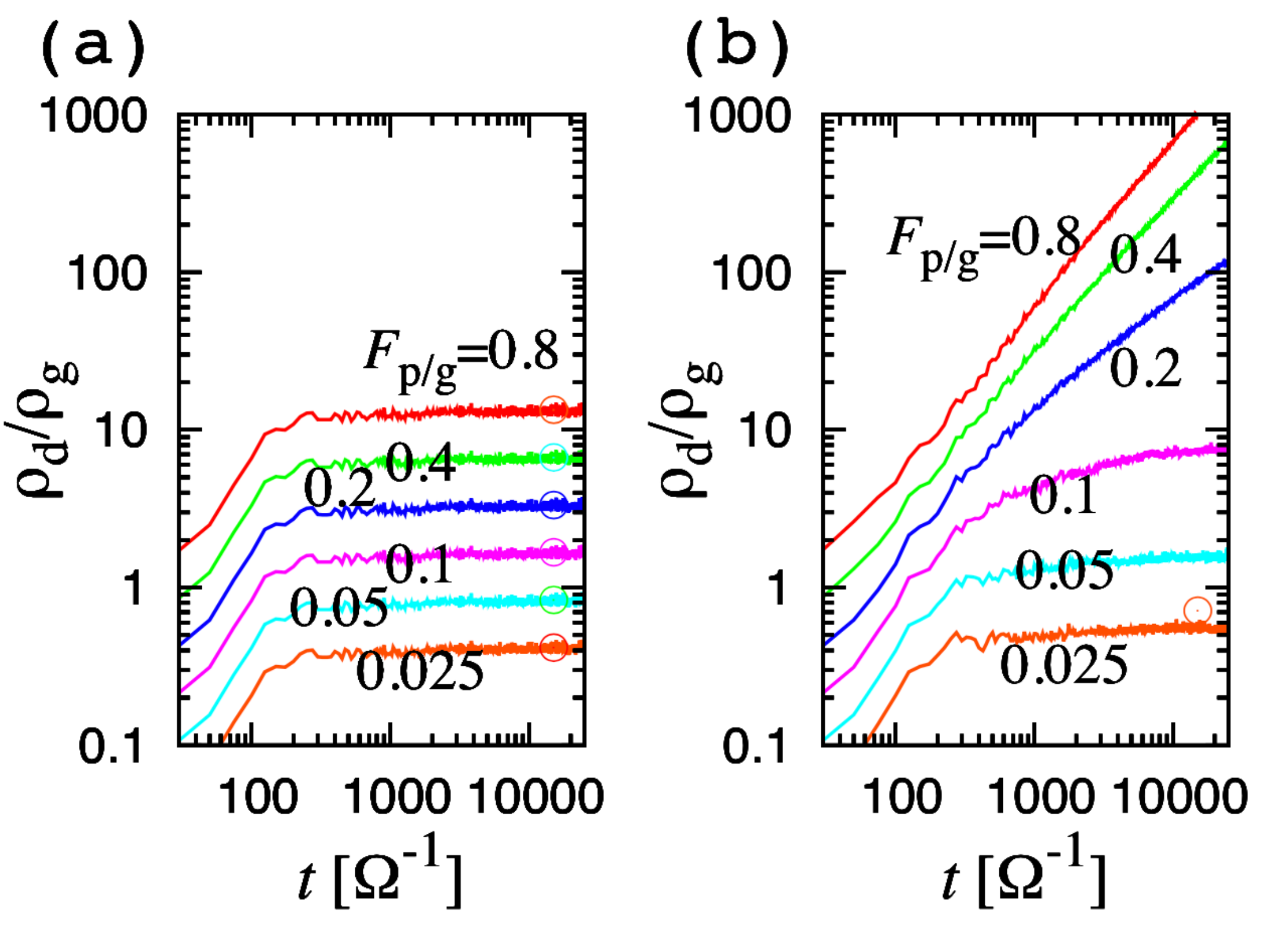}
\caption{   
The time evolution of the maximum
$\rhod/\rhog$.
The back-reaction to $\vr$ is included in the results in panel (b), but 
not in panel (b).
The light blue, magenta, blue, green, and red curves are
the results with $F_{\rm p/g}=0.025, 0.05, 0.1, 0.2, 0.4$ and 0.8, respectively.
The other parameters are fixed as $\acc=10^{-2}$ and 
$\Delta x_{\rm subl} = 0.1\hg$ and $\Delta z_{\rm subl} = 0.03\hg$.
To mimic the settings of \citet{Ida_Guillot2016}, we artificailly set
$\Dr=0$, $\Dz=10^{-5}$ and $\st = 10^{-2}$
(however, the relatively large $\st$ for the silicate dust is not
reflected to the dust drift speed).
The latter two parameters are to maintain $\hd = \Delta z_{\rm subl}$.
The circles in panel (a) represent the analytical solution 
given by Eq.~(\ref{eq:SO19}).
That in panel (b) is the analytical solution 
for non-divergent case ($\fpg=0.025$) given by Eq.~(\ref{eq:IG3}).
}
\label{fig:IG}
\end{figure}

\citet{Ida_Guillot2016} adopted the following setup:
\begin{itemize}
\item The turbulent diffusion is neglected for simplicity.
\item The silicate particle pile-up is fast enough
that $\hd$ is not increased from $\Delta z_{\rm subl} (= \hp)$ by the vertical stirring.
\end{itemize}
To mimic the \citet{Ida_Guillot2016}'s setup, we set
the following in the Monte Carlo simulations:
\begin{itemize}
\item We adopt $\Dr=0$, $K=0$, and $\Delta z_{\rm subl} = 0.03 \hg$.
\item To keep $\hd = \Delta z_{\rm subl}$ as an equilibrium,
we adopt $\Dz=10^{-5}$ and $\st = 10^{-2}$.
\item The dust particle drift speed is set to be $\vr = - \Lambda u_\nu$,
to which the artificially enlarged $\st$ is not reflected (Eq.~(\ref{eq:vr})).
For comparison, $\vr = - u_\nu$ (without the back-reaction to $\vr$)
is also examined.
For gas, $u_r = -u_\nu$ is always assumed.
\end{itemize}
In the later results, we always set $\stpeb = 10^{-1}$ and $\st = 10^{-5}$
and consistently calculate $\hp (= \Delta z_{\rm subl})$ by Eq.~(\ref{eq:H_peb}) 
and the evolution of $\hd$ from $\Delta z_{\rm subl}$ 
by the vertical and radial diffusion, with given $\Dz$ and $\Dr$.
The above artificial setting is used only in the particular runs in Figure~\ref{fig:IG}.

Figure~\ref{fig:IG} shows the time evolution of 
the maximum value of $\rhod/\rhog$ 
(the value on the midplane near the snow line) for different values of $F_{\rm p/g}$,
in the case of $\acc=10^{-2}$, $\Delta x_{\rm subl} = 0.1\hg$, and $\Delta z_{\rm subl} = 0.03 \hg$. 
In panel (a), the back-reaction to $\vr$ is not included ($\Lambda=1$) in the Monte Carlo calculations,
that is, $\vr = u_r = - u_\nu$,
which corresponds to the result in \citet{Schoonenberg2017} except that they assumed $\hdg=1$.
The maximum $\rhod/\rhog$ quickly reaches the equilibrium
values that are given by
Eq.~(\ref{eq:Z_Sigma}) with $\vr = u_r = - u_\nu$ and $Z_\Sigma = Z \, \hdg$ as
\begin{equation}
\frac{\rhod}{\rhog} \simeq f_{\rm d/p} h_{\rm d/g}^{-1} F_{\rm p/g}
\simeq 16.7 \, F_{\rm p/g}.
\label{eq:SO19}
\end{equation}
The open circles in panel (a) represent this analytical solution
for each of $\fpg$, which completely agree with the numerical results here.

In Figure~\ref{fig:IG}b, the back-reaction to $\vr$ is included ($\Lambda < $),
corresponding to \citet{Ida_Guillot2016}'s setup.
They predicted that runaway pile-up occurs for $F_{\rm p/g} > h_{\rm d/g}  f_{\rm d/p}^{-1}$,
which is $F_{\rm p/g} > 0.06$ for the parameters adopted here.
This panel shows a clear runaway pile-up for $\fpg \ga 0.1$,
which is consistent with the prediction.

Comparison between Figure~\ref{fig:IG}a and b clearly shows that
the back-reaction to $\vr$ of the silicate particles
plays an essential role in the occurrence of the runaway pile-up.
The increase in the pile-up slows down the drift velocity and
accordingly increases the pile-up itself.
Because it is certain that the back-reaction to $\vr$, we include it in the results 
in the rest of the paper.

If the supply rate of silicate particles ($f_{\rm d/p} \dot{M}_{\rm p}$)
exceeds a threshold value, the mode of the pile-up becomes runaway.
The threshold value must be regulated by the local value of $\rhod/\rhog$,
because the back-reaction becomes effective once $\rhod/\rhog$ exceeds unity.
While $\dot{M}_{\rm p}$, more exactly $F_{\rm p/g}$,
determines $Z_\Sigma = \sigd/\sigg$, the local value of $Z=\rhod/\rhog$ is
regulated by $\hd$ and $Z_\Sigma$ ($Z=Z_\Sigma/\hdg$) near the snow line.
The silicate dust particle scale height $\hd$ is regulated by
the pebble scale height $H_{\rm p}$ ($=\Delta z_{\rm subl}$ in our simulation)
and turbulent mixing parmeters, $\Dz$ and $\Dr$, as shown in Section 4.3.

\subsection{Typical results}

\begin{figure*}
\includegraphics[width=16cm]{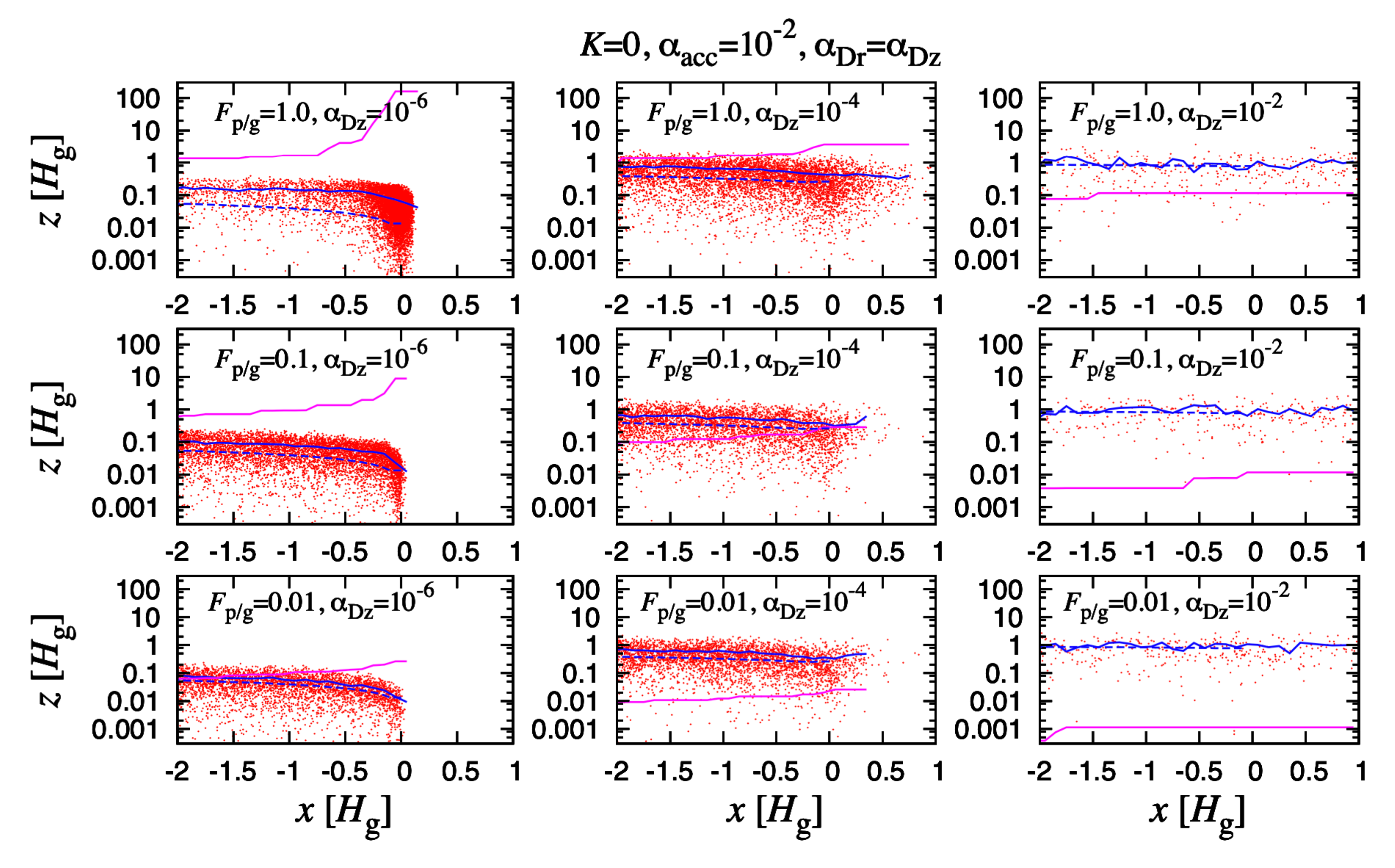}
\caption{  
Snapshots of silicate particles at $t=10^4\Omega^{-1}$ (red dots),
and the root mean square of $z$ (blue solid curves) at each grid of $x$. 
The dashed blue curves are given by Eq.~(\ref{eq:hdg_star}) 
with $\Delta x_{\rm subl}$ replaced by $\max(\Delta x_{\rm subl}, | x |)$ (see the discussion at the end of Section 4.3.2).
The magenta curve represents $\log_{10}(\rhod/\rhog)$ near the midplane.
We set $\acc = 10^{-2}$, $\st = 10^{-5}$, and $f_{\rm d/p}=0.5$, and $\mdotg=10^{-8}M_\odot/{\rm yr}$. 
The nine panels adopt different values of $F_{\rm p/g}$ and $\Dz=\Dr$, as indicated in each panel. 
}
\label{fig:K=0}
\end{figure*}

\begin{figure*}
\includegraphics[width=16cm]{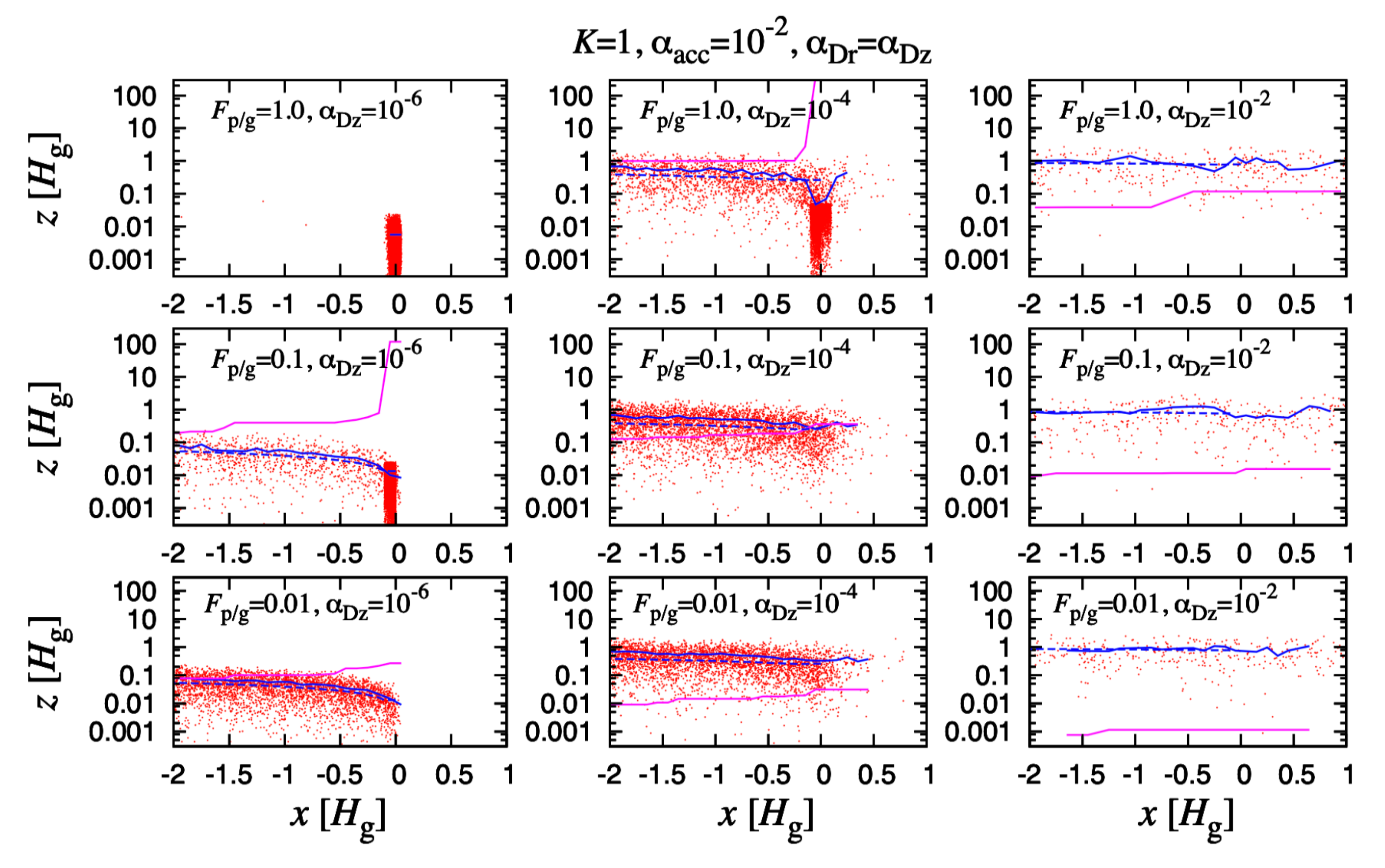}
\caption{  
The same as Fig.~\ref{fig:K=0} except for $K=1$.
}
\label{fig:K=1}
\end{figure*}

We first show typical results in our simulations for different
$\fpg$ and $\Dz \,(=\Dr)$.
As we will show below, the radial diffusion also contributes to $\hd$.
The back-reaction to $\vr$ is included.
In these results, $\Delta x_{\rm subl} = 0.1 \hg$ is adopted, as well as $\acc=10^{-2}$.

Figure~\ref{fig:K=0} show the snapshots of silicate particles at $10^4\Omega^{-1}$
for the cases without the
back-reaction to diffusion coefficients ($K=0$). 
In the right panels with $\Dz=\Dr=10^{-2}$ ($\Dz /\st  = 10^3$), 
the vertical mixing is fast enough to
realize the upper limit of $\hd$ ($\hd \simeq \hg$)
for all of $\fpg = 0.1, 0.3$ and 1.
Even for high mass flux of pebbles ($\fpg = 1.0$),
no pile-up of the silicate particles is found for $\Dz=\Dr=10^{-2}$. 
In the case of $\Dz=\Dr=10^{-6}$, on the other hand,
the mixing is so weak that $\hd$ is not increased to
the equilibrium value [$\simeq (\Dr/\st)^{1/2}\hg \sim 0.3\hg$]
in the range of $x > -2 \,\hg$.
Nevertheless, $\hd$ near the injection point ($x\sim 0$)
is much higher than $\Delta z_{\rm subl} = \hp = 0.003 \hg$.
This means that $\hd$ is regulated by $\Dr$ as well as by $\Dz$,
which will be discussed in Section 4.3.

Figure~\ref{fig:K=1} show the results with the back-reaction to diffusion coefficients ($K=1$). 
In general, the runaway pile-up is more pronounced for smaller $\Dz=\Dr$ and larger $\fpg$.
The magenta curves represent $\log_{10}(\rhod/\rhog)$ near the midplane.
The midplane $\rhod/\rhog$ usually takes the maximum value, 
at the injection (sublimation) region, $x \sim 0$, which is denoted by $(\rhod/\rhog)_0$.
In the panels with $(\rhod/\rhog)_0 \la 1$
(the panels with $\Dz=10^{-2}$ and those with $\Dz=10^{-4}$ and $\fpg=0.3$ and 0.1),
the results are similar to those in Figure~\ref{fig:K=0}. 
If $\rhod/\rhog \ga {\rm several}$ in Figure~\ref{fig:K=0}, however,
the pile-up is much more pronounced in this case, because   
the radial and vertical diffusions become much weaker
as the pile-up proceeds.

\begin{figure}[htb]
\includegraphics[width=90mm,angle=0]{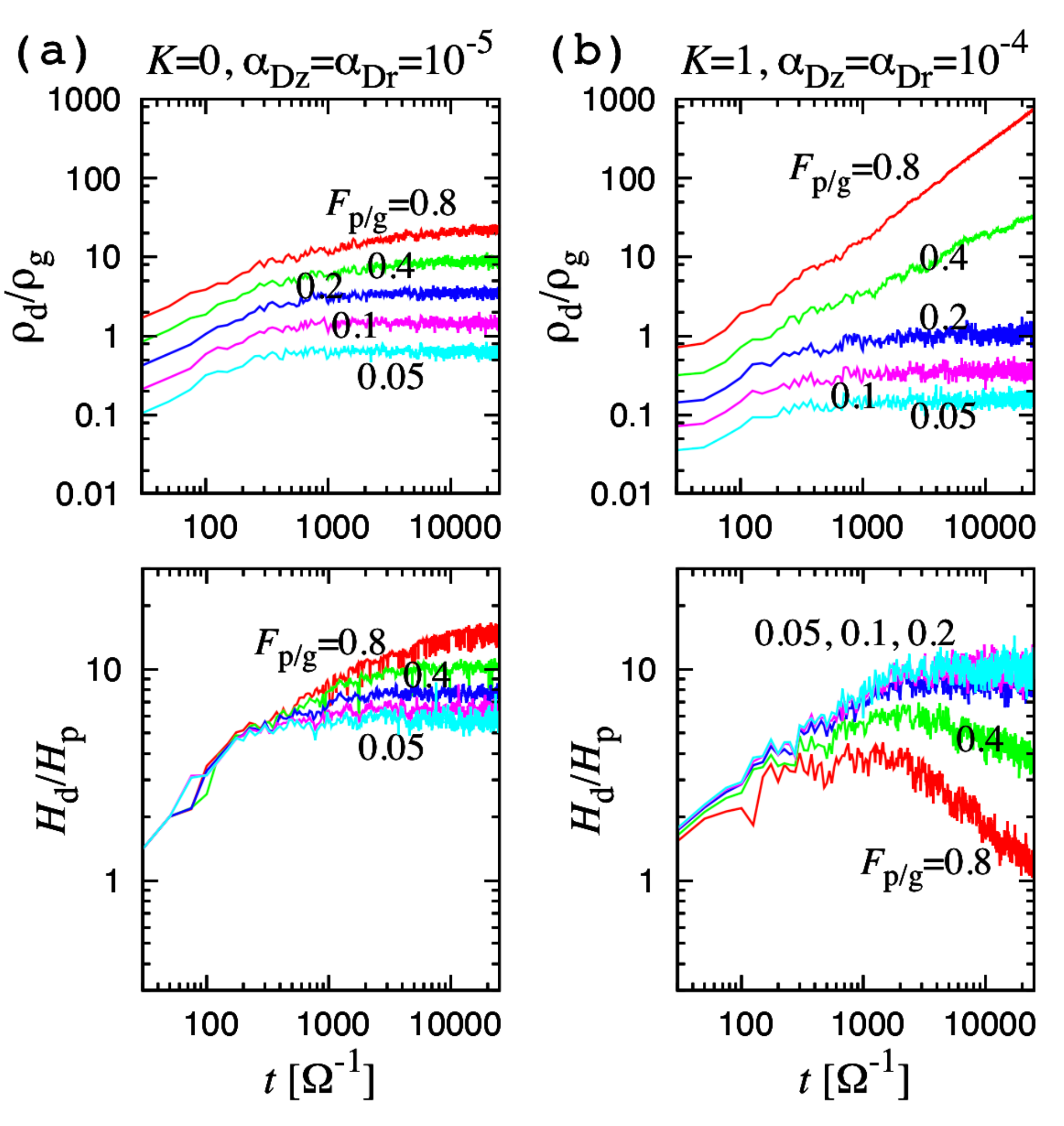}
\caption{The time evolution of $(\rhod/\rhog)_0$
and the dust scale height at the maximum
$\rhod/\rhog$ locations scaled by the initial one ($=\hp$).
The back-reaction to diffusion is included with $K=1$
in panel (b), but not ($K=0$) in panel (a),
while the back-reaction to $\vr$ is included in both panel (a) and (b).
The light blue, magenta, blue, green, and red curves are
the results with $F_{\rm p/g}=0.05, 0.1, 0.2, 0.4$ and 0.8, respectively.
The other parameters are fixed as $\acc=10^{-2}$, 
$\Delta x_{\rm subl} = 0.1$, and $\Delta z_{\rm subl} = 0.03$.
The diffusion parameters are $\Dz=\Dr=10^{-5}$ in panel (a) and $\Dz=\Dr=10^{-4}$
in panel (b). 
}
\label{fig:tevol}
\end{figure}

Figure~\ref{fig:tevol} show the time evolution of $(\rhod/\rhog)_0$ and $\hd/\hp$ at the maximum
$\rhod/\rhog$ locations ($h_{\rm d/g,0}$), for different $\fpg$.
In panel (a), the back-reaction to diffusion is not included ($K=0$).
As Figures~\ref{fig:K=0} and \ref{fig:K=1} suggest,
the transient $\Dz$ $(=\Dr)$ values between the runaway pile-up and non-pile-up cases 
are $\Dz \simeq 10^{-5}$ for $K=0$ and
$\Dz \simeq 10^{-4}$ for $K=1$, so that we plot the results with
$\Dz = 10^{-5}$ in panel (a) and $\Dz = 10^{-4}$ in panel (b).
These figures show that transition to the runaway pile-up is
much clearer in the case of $K=1$.

This difference comes from the $\hd$ evolution.
The comparison with the result of
\citet{Ida_Guillot2016} in Section 2 suggests that
$\hd$ is directly related to a threshold value of $\fpg$ for the runaway pile-up:
A lower $\hd$ leads to a lower threshold value of $\fpg$.
However, Figure~\ref{fig:tevol} show 
that $\hd$ increases as $\rhod/\rhog$ increases in the case of $K=0$, 
while it decreases for $K=1$.
Therefore, the pile-up suffers a negative feedback for $K=0$,
but it suffers a positive feedback, resulting in much clearer transition
to the runaway pile-up for $K=1$. 

In both cases, the back-reaction to $\vr$ is included.
As the pile-up proceeds, $\vr$ decreases, which
enhances the pile-up.
However, we also include the vertical stirring.
As $\vr$ decreases, the silicate particles are stirred up 
for longer time before they leave the sublimation region,
resulting in higher $\hd$ near the sublimation region.
In panel (a), $\hd$ near the sublimation region quickly asymptotes to an equilibrium value
 for $\fpg \le 0.2$ (lower $(\rhod/\rhog)_0$ cases);
it increases with time for $\fpg = 0.8$ (the runaway pile-up cases).
In panel (b), in the runaway pile-up cases ($\fpg =0.4$ and 0.8),
$\hd$ initially increases with time.
However, after $\rhod/\rhog$ exceeds $\sim 1$, 
the diffusion coefficients becomes so small that
 the settling of the particles overwhelms the vertical stirring and $\hd$ decreases.   
Thus, the different responses to the pile-up between $K=0$ and $K=1$
result in different $\hd$ and $\rhod/\rhog$ evolution in the pile-up cases.
As we discussed in Section 3.2, $K=1$ (with the diffusion back-reaction) is more realistic.

\subsection{Silicate particle scale height}

\subsubsection{Monte Carlo simulation results}

\begin{figure}[htb]
\includegraphics[width=90mm,angle=0]{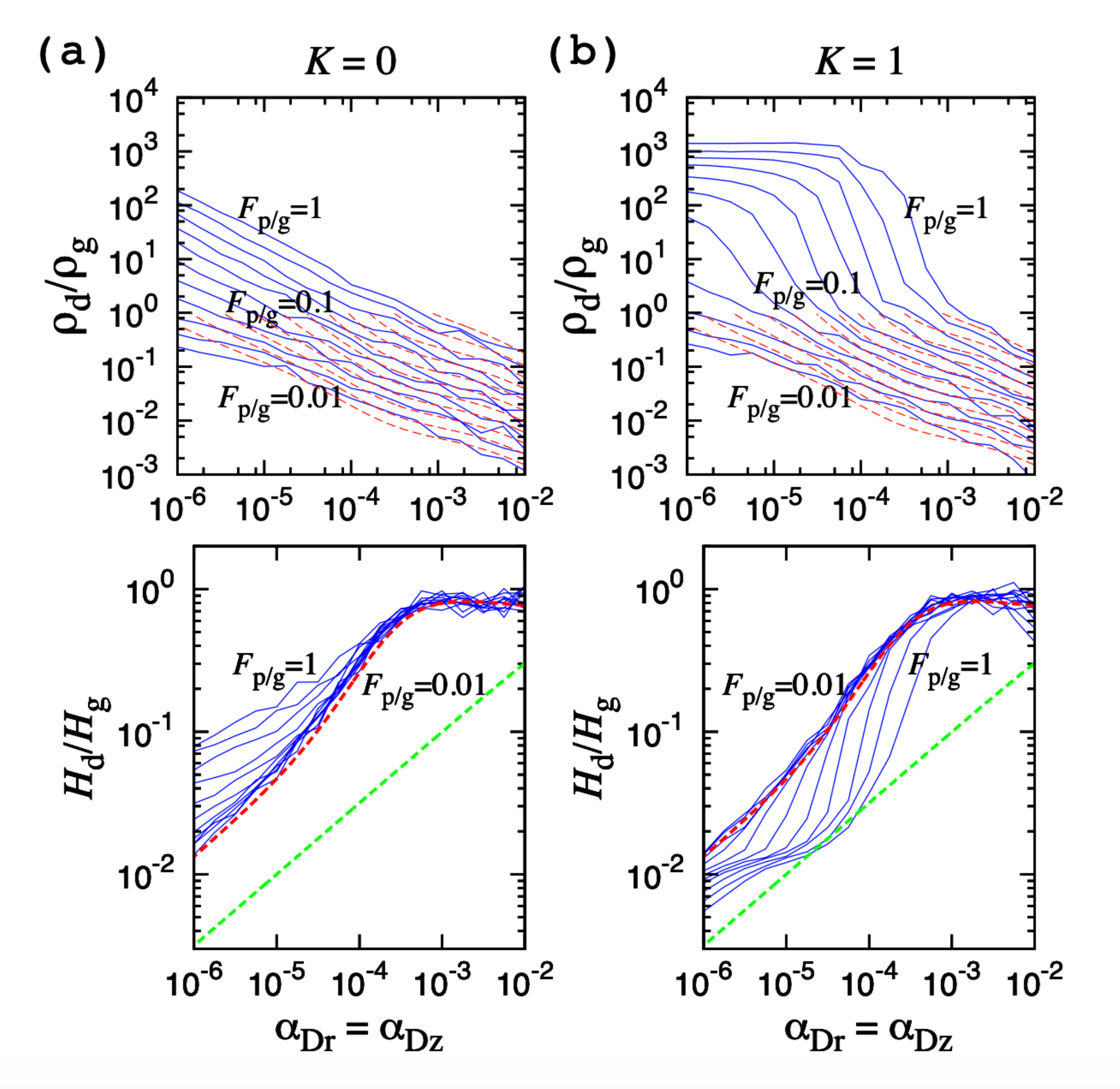}
\caption{The maximum
$\rhod/\rhog$ and $\hd/\hg$ at the maximum $\rhod/\rhog$ locations
at $t = 15000 \Omega^{-1}$ as a function of $\Dz=\Dr$
for different $\fpg$.
The back-reaction to diffusion is included with $K=1$
in panel (b), but not ($K=0$) in panel (a).
The other parameters are fixed as $\acc=10^{-2}$, 
$\Delta x_{\rm subl} = 0.1 \hg$, and $\Delta z_{\rm subl}\hp$ given by Eq.~(\ref{eq:H_peb}).
In the upper and bottom panels, the red dashed curves are the analytical formulas given by 
Eqs.~(\ref{eq:rho_anly_diff}) and (\ref{eq:hdg_star}) with Eq.~(\ref{eq:hdg}).
In the lower panels, the green dashed lines represents $\Delta z_{\rm subl}/\hg$.
}
\label{fig:hd_anly}
\end{figure}

Figure~\ref{fig:hd_anly} show $(\rhod/\rhog)_0$ and $\h0$ at $t = 15000 \Omega^{-1}$
as a function of $\Dz=\Dr$ for different $\fpg$.
This figure more clearly shows that
the asymptotic equilibrium values of
$\h0$ are independent of $\fpg$ in the non runaway pile-up cases
which is suggested by Figure~\ref{fig:tevol}.
The envelope curves in the plots for (a) $K=0$ and (b) $K=1$  
match the analytical formula
derived in the next subsubsection 
(Eqs.~(\ref{eq:rho_anly_diff}) and (\ref{eq:hdg_star}) with Eq.~(\ref{eq:hdg})).
The equilibrium values are generally larger than the scale height
of the injected silicate particles $\Delta z_{\rm subl}$, due to
the effects of vertical and radial diffusion.
In panels (a) and (b), the values of $\h0$ that deviate from the envelope curves
correspond to the runaway pile-up; $\h0$ deviates to higher values for $K=0$
and lower values for $K=1$.
If we calculate on longer timescales, the deviations increase.

\subsubsection{Analytical formula for the silicate scale height}

The red dashed curves in Figure~\ref{fig:hd_anly} are analytical formulas for the non pile-up cases,
which is derived as follows.
The timescale to drift by $\Delta x_{\rm subl}$ is
\begin{align}
t_{\rm drift} & \simeq \frac{\Delta x_{\rm subl}}{\vr} \nonumber \\
& \simeq \frac{\Delta x_{\rm subl}}{(3/2) \Lambda \acc (H_{\rm g}/r)^2 \vk}
\simeq \frac{\Delta x_{\rm subl}/\hg}{(3/2) \Lambda \acc (H_{\rm g}/r) \Omega}.
\end{align}
As the pile-up proceeds ($\Lambda = 1/(1+Z)$ decreases), the particle drift becomes slower
and accordingly $t_{\rm drift}$ becomes longer. 
The vertical diffusion length 
during $t_{\rm drift}$ is
\begin{align}
z_{\rm diff} \simeq \sqrt{D_z t_{\rm drift}}
& \simeq \left( \frac{2}{3} \alpha_{D,z}\Lambda^{K} \hg^2 \Omega \times  \frac{\Delta x_{\rm subl}/\hg}{\Lambda\acc (H_{\rm g}/r) \Omega}  \right)^{1/2} \nonumber\\
 & \simeq \left( \frac{2}{3} \Lambda^{K-1}
 \frac{\Delta x_{\rm subl}/\hg}{H_{\rm g}/r} \frac{\alpha_{D,z}}{\acc} \right)^{1/2}\hg.
\end{align}
This expression suggests that $z_{\rm diff}$ becomes higher
as the pile-up proceeds ($\Lambda$ decreases) for the case of $K=0$.
Hereafter we derive the equilibrium value of $\hd/\hg$ for the non pile-up cases
with $\Lambda \simeq 1$ (the envelope curves in Figure~\ref{fig:hd_anly}). 
If $Z\gg 1$ for the equilibrium value of $\hd/\hg$, 
the runaway pile-up occurs,
so that $Z$ keeps increasing and the equilibrium value of $\hd/\hg$ no more exists.

If only vertical diffusion is considered,
the silicate particle scale height scaled by $\hg$ is predicted as
\begin{align} 
h_{\rm d/g} = \frac{\hd}{\hg} & \simeq \frac{\max \left( \Delta z_{\rm subl}, z_{\rm diff} \right)}{\hg} 
\simeq \frac{(  \Delta z_{\rm subl}^2 + z_{\rm diff}^2 )^{1/2}}{\hg} \nonumber \\
 & \simeq 
 \left(\frac{\Delta z_{\rm subl}^2}{\hg^2} + \frac{2}{3} \frac{\Dz}{\acc} 
\frac{\Delta x_{\rm subl}/\hg}{\hg/r}\right)^{1/2}.
\end{align}

Furthermore, during $t_{\rm drift}$,
the particles are radially mixed in the  range of 
$x_{\rm diff} \simeq \sqrt{D_r t_{\rm drift}}
\sim (\Dr/\Dz)^{1/2} z_{\rm diff}$.
In general, the dust scale height increases with the distance from
the snow line.
If $x_{\rm diff} \gg \Delta x_{\rm subl}$, the radial diffusion 
brings the high $z$ particles back to the injection region 
and raises the value of $h_{\rm d/g}$ there.
To take this effect into account,
we found through 
comparison with the Monte Carlo simulation results
that a good-fit analytical formula is obtained by 
multiplying the above $h_{{\rm d/g}}$ by
the factor of $[1 + (x_{\rm diff}/\Delta x_{\rm subl})^2]$.
When $\Dr$ is comparable to $\acc$, 
the silicate scale height is radially smoothed out and
the effect of radial mixing for the scale height becomes weak.
To take this into account, 
we replace $\Dr/\acc$ in $(x_{\rm diff}/\Delta x_{\rm subl})^2$
by $(\Dr/ \acc)/[1+ (C_{r, \rm diff} \, \Dr/\acc )^2]$ in the above equation.
Comparing the formula with the numerical results
(Figs.~\ref{fig:hd_anly} and \ref{fig:hd_anly4}), we empirically adopt $C_{r, \rm diff}=10$.
With these modifications, 
\begin{align}
h_{\rm d/g,*} \simeq & 
 \left( \frac{\Delta z_{\rm subl}^2}{\hg^2} + \frac{2}{3} \frac{\Dz}{\acc} 
\frac{\Delta x_{\rm subl}/\hg}{\hg/r}\right)^{1/2}
\nonumber\\
  & \times \left[1 + \frac{2}{3}
  \frac{(\Dr/\acc)}{1+ (C_{r, \rm diff}\,\Dr/\,\acc)^2} \frac{1}{(\hg/r)(\Delta x_{\rm subl}/\hg)} \right].
\label{eq:hdg_star}
\end{align}
Because $h_{\rm d/g}$ cannot exceed the equilibrium value 
$h_{\rm d/g,eq} \simeq (1+\st/\Dz)^{-1/2} $,
the final form is
\begin{equation}
h_{\rm d/g} \simeq (h_{\rm d/g,eq}^{-1} + h_{\rm d/g,*}^{-1})^{-1}.
\label{eq:hdg}
\end{equation}
As the lower panels in Figure~\ref{fig:hd_anly} show,
the analytical formula reproduces the envelope curves of $\hd/\hp$ obtained by 
the numerical simulations for the non pile-up cases.
We note that $\h0$ and $(\rhod/\rhog)_0$  
depend on the ratios of $\Dr/\acc$ and $\Dz/\acc$, but not on
their absolute values.

Equation~(\ref{eq:hdg_star}) is the analytical formula for $\hdg$ at $x \sim 0$.
This formula can be extrapolated to any $x$ by replacing
$\Delta x_{\rm subl}$ with $\max(\Delta x_{\rm subl}, | x |, \epsilon)$,
where $\epsilon$ is added to avoid the divergence
in the case of extremely small $\Delta x_{\rm subl}$; we adopt $\epsilon = 0.01 \, \hg$.
In Figs.~\ref{fig:K=0} and \ref{fig:K=1},
the scale height distribution obtained by the Monte Carlo simulations are well reproduced by the
analytical formula (the blue dashed curves) except for the runaway pile-up cases where
the back-reaction modulates the scale height as we discussed in this subsection.

\subsection{Silicate particle pile-up}
\subsubsection{Analytical formula for the silicate pile-up}

We can also derive an analytical formula for $(\rhod/\rhog)_0$
for the non runaway pile-up cases and for the phase before the runaway pile-up proceeds.
If the estimated $(\rhod/\rhog)_0$ exceeds $\sim 1$,
the actual value of $\rhod/\rhog$ 
should deviate from the estimate and increase with time.

\begin{figure}[htb]
\includegraphics[width=95mm,angle=0]{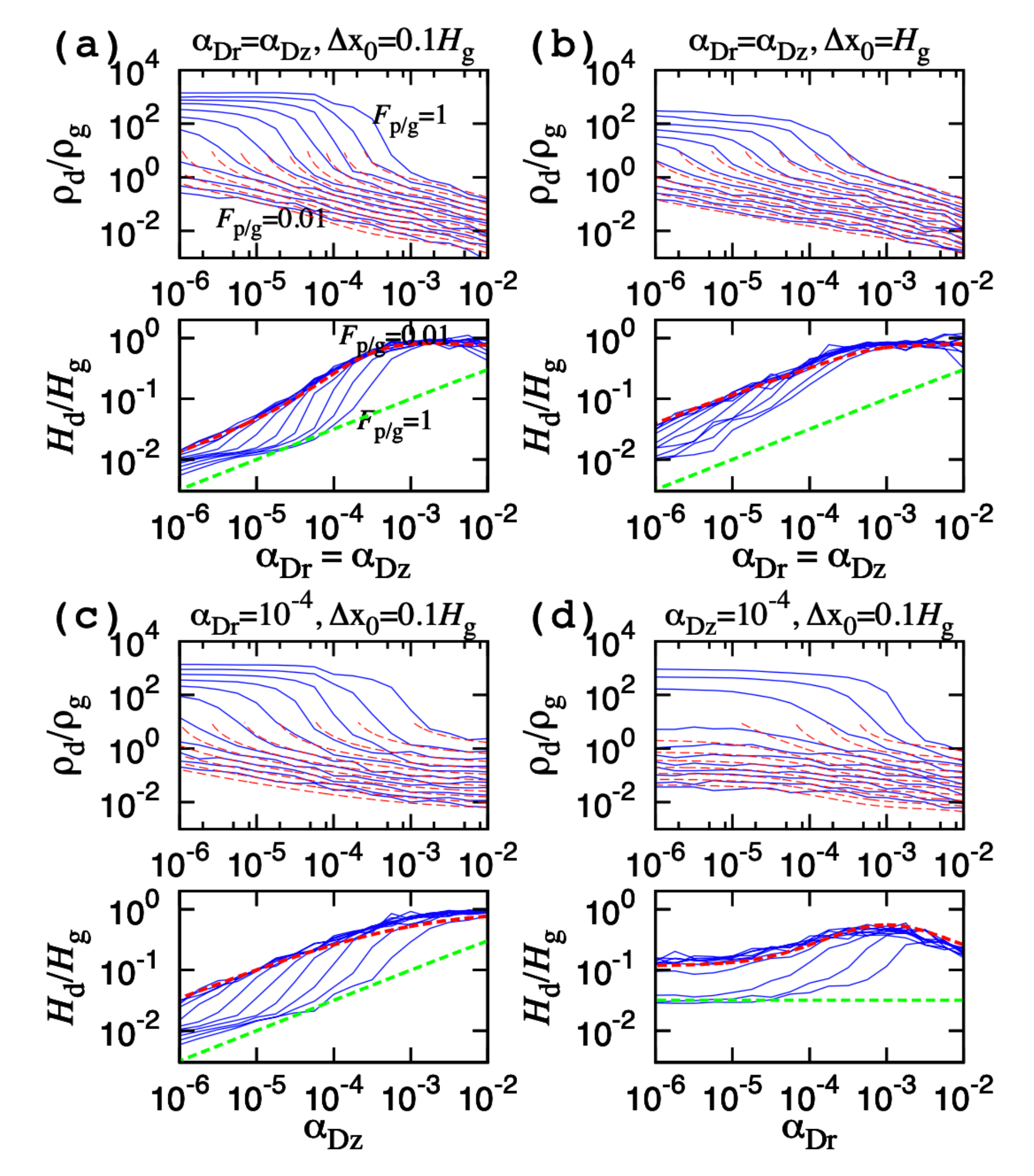}
\caption{Same as Figure~\ref{fig:hd_anly} except for
$\Dz/\Dr$ relations and the value of $\Delta x_{\rm subl}$.
The maximum $\rhod/\rhog$ and $\hdg$ $(= \hd/\hg)$ there for different $\fpg$
are given as functions of
(a) $\Dr = \Dz$ with $\Delta x_{\rm subl} = 0.1\, \hg$,
which is the same as Figure~\ref{fig:hd_anly}b for a comparison,
(b) $\Dr = \Dz$ with $\Delta x_{\rm subl} =\hg$,
(c) $\Dz$ with $\Dr=10^{-4}$ and $\Delta x_{\rm subl} =0.1\, \hg$,
and (d) $\Dr$ with $\Dz=10^{-4}$ and $\Delta x_{\rm subl} =0.1\, \hg$.
For all the cases, $K=1$ is adopted.
In the upper and bottom panels, the red dashed curves are the analytical formulas given by 
Eqs.~(\ref{eq:hdg_star}) and (\ref{eq:rho_anly_diff}) with corresponding parameter values.
}
\label{fig:hd_anly4}
\end{figure}

Basically, $(\rhod/\rhog)_0$ is predicted by Eq.~(\ref{eq:IG3})
with the scale hight formula, Eqs.~(\ref{eq:hdg}) and (\ref{eq:hdg_star}).
A possible detailed correction is to include the effect of
outward diffusion flux ($F_{\rm d,D}$) of silicate particles beyond the snow line
that can be found in the snapshots with $\Dr=\acc=10^{-2}$
in Figs.~\ref{fig:K=0} and \ref{fig:K=1}.
As shown in Appendix B, calibrating with the results of Monte Carlo simulations
in Figs.~\ref{fig:hd_anly} and \ref{fig:hd_anly4},
the effect can be done 
with $f_{\rm d/p} F_{\rm p/g}$ replaced by  
$F_{\rm d,net} = f_{\rm d/p} F_{\rm p/g} - F_{\rm d,D}$,
where 
\begin{equation}
F_{\rm d,net} = f_{\rm d/p} F_{\rm p/g}
\left[1 - \left(1+ \frac{15}{2}\frac{\hg}{r}\frac{\acc}{\Dr}\right)^{-1} \right].
\label{eq:rho_anly_diff6}
\end{equation}
Accordingly, Eq.~(\ref{eq:IG3}) is reduced to
\begin{eqnarray}
\left(\frac{\rho_{\rm d}}{\rho_{\rm g}}\right)_0 =
\frac{F_{\rm d,net}}{h_{\rm d/g}  - F_{\rm d,net}}.
\label{eq:rho_anly_diff5}
\end{eqnarray}
This formula reproduces the numerical results 
in Figs.~\ref{fig:hd_anly} and \ref{fig:hd_anly4}, as long as $\rhod/\rhog \la 1$.
Note, however, that if the effect of sticking to icy pebbles
by silicate particles that diffuse beyond the snow line is considered,
most of the silicate grains beyond the snow line
may eventually come back with the pebbles,
so that $F_{\rm d,net}$ would become more similar to $f_{\rm d/p} F_{\rm p/g}$.

We have derived
the analytical formulas for the silicate particle scale height,
Eqs.~(\ref{eq:hdg_star}) and  (\ref{eq:hdg}), and
those for the pile-up, Eqs.~(\ref{eq:rho_anly_diff6}) and (\ref{eq:rho_anly_diff5}).
The formulas make clear the intrinsic physics
of the pile-up of the silicate particles released from
the drifting icy pebble by their sublimation, which is a complicated process.
Here we study the parameter dependences using the analytical formulas 
and the simulations with $K=1$.

The formulas imply that the results are scaled by
$\Dz/\acc$ and $\Dr/\acc$, for given $\Delta z_{\rm subl}$,
$\Delta x_{\rm subl}$, and $\hg/r$.
We also carried out the Monte Carlo simulations with 
$\acc = 10^{-3}$ and confirmed that 
the plots in Figure~\ref{fig:hd_anly} are almost the same
as long as we use $\Dr/\acc$ ($=\Dz/\acc$) as the horizontal coordinate.

The sublimation width, $\Delta x_{\rm subl}$, is an independent parameter.
The results with $\Delta x_{\rm subl}=\hg$ (Fig.~\ref{fig:hd_anly4}b), 
are compared with those with $\Delta x_{\rm subl}=0.1\hg$ (Fig.~\ref{fig:hd_anly4}a,
which are identical to Fig.~\ref{fig:hd_anly}b).
In Fig.~\ref{fig:hd_anly4}b, 
the injected particles are distributed in a 10 times broader region
than in the case of Fig.~\ref{fig:hd_anly4}a.
As a result, the vertical mixing is more effective ($t_{\rm drift}$ is longer)
and the pile-up (the maximum $\rhod/\rhog$) is lower.

So far, the results with $\Dr=\Dz$ have been shown.
In Figure~\ref{fig:hd_anly4}c,
$\Dr$ is fixed and $\Dz$ is independently changed.
In Figure~\ref{fig:hd_anly4}d, $\Dz$ is fixed.
The envelope curves for $\hd/\hg$ ($=\hdg$) are reproduced by the analytical formula
also in these cases,
which strongly suggests that the correction factors with the vertical and radial diffusion
are physically correct.

\subsubsection{The condition for the silicate runaway pile-up}
 
\begin{figure*}
\begin{center}
\includegraphics[width=14cm]{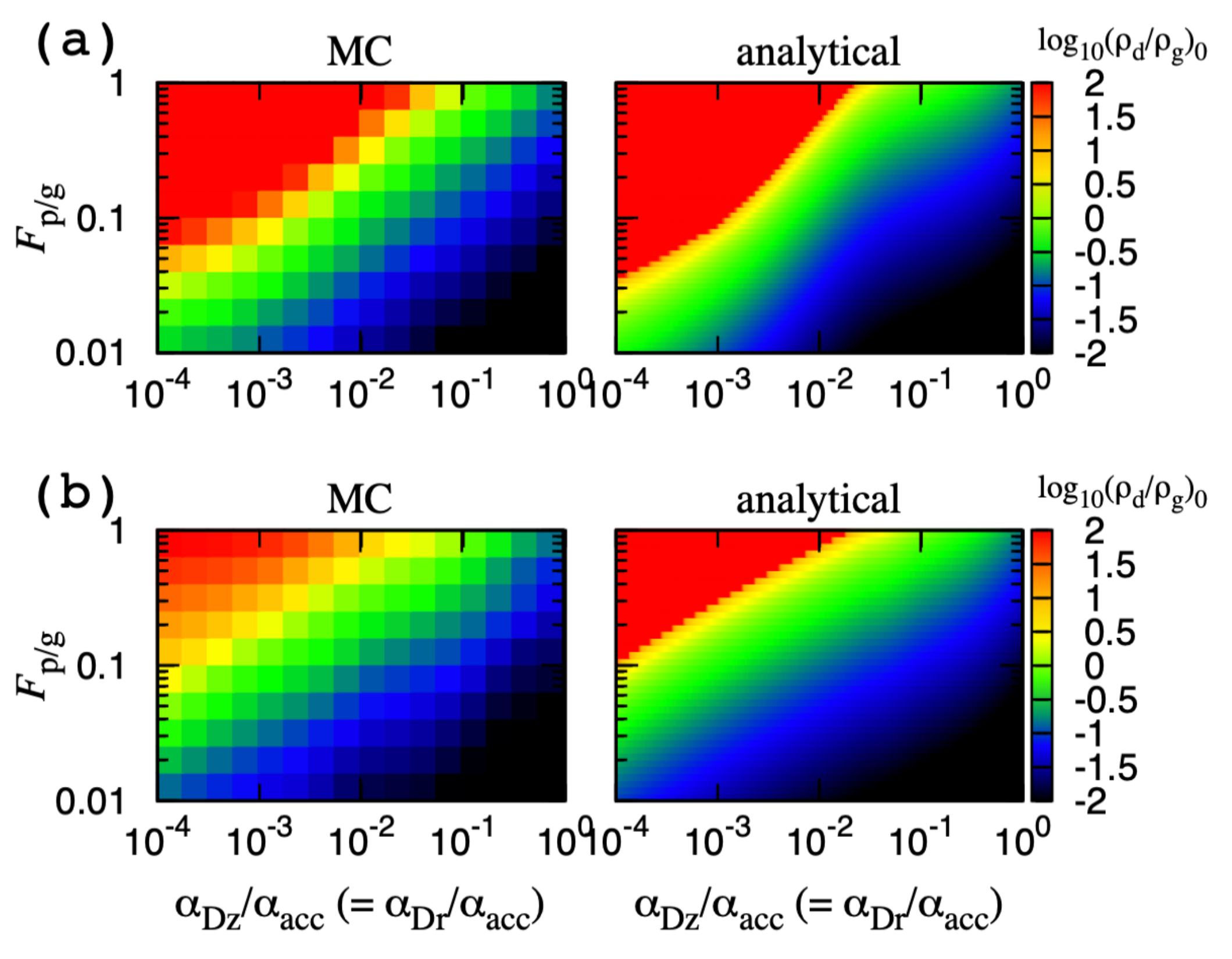}
\caption{The contour maps of the maximum
$\rhod/\rhog$ on the plane of $\Dz/\acc (=\Dr/\acc)$ and $\fpg$
with (a) $\Delta x_{\rm subl} = 0.1 \, \hg$, 
and (b) $\Delta x_{\rm subl}$ given by Eq.~(\ref{eq:dx0a2}).
In both cases, $\acc=10^{-2}$ and $K=1$ are adopted.
The left panels are the results of Monte Carlo simulations
at $t = 15000 \Omega^{-1}$ 
and the right ones are the results using analytical formulas
given by Eq.~(\ref{eq:rho_anly_diff}) with
Eqs.~(\ref{eq:hdg_star}), (\ref{eq:hdg}), (\ref{eq:rho_anly_diff2}),
and the corresponding $\Delta x_{\rm subl}$.
The Monte Carlo simulations were performed
with a grid size of $\Delta \log_{10} (\Dz/\acc)=\Delta \log_{10} \fpg = 0.2$.
We used more refined grids to make the maps by the analytical formulas. 
The color scales are based on $\log_{10} (\rhod/\rhog)$.
}
\label{fig:map}
\end{center}
\end{figure*}

The analytical formulas reproduce the numerical results for broad range of multi-dimensional parameters. 
Figure~\ref{fig:map}a shows 
the contour maps of the maximum
$\rhod/\rhog$ on the plane of $\Dz/\acc (=\Dr/\acc)$ and $\fpg$ at $t = 1.5\times 10^4 \Omega^{-1}$ for 
$\Delta x_{\rm subl} = 0.1\,\hg$.
In Figure~\ref{fig:map}b, 
$\Delta x_{\rm subl}$ is given by Eq.~(\ref{eq:dx0a2}).
In both cases, $\acc=10^{-2}$ and $K=1$.
The left and right panels are the results of Monte Carlo simulations
and the analytical results, respectively.
The red regions represent the parameter ranges of runaway pile-up of silicate particles.
As expected, the analytical formulas reproduce the numerical results. 

For these simulations, $\acc=10^{-2}$ and 
$\Dz (= \Dr)$ is in the range of $[10^{-6},10^{-2}]$.
The analytical formula to predict $(\rhod/\rhog)_0$ 
is Eq.~(\ref{eq:rho_anly_diff6}) with Eqs.~(\ref{eq:rho_anly_diff5}),
(\ref{eq:hdg_star}), and (\ref{eq:hdg}).
They are functions of  $\Dr/\acc$ and $\Dz/\acc$ for given $\Delta x_{\rm subl}$,
as long as the term $\Delta z_{\rm subl}^2$ in Eq.~(\ref{eq:hdg_star})
is negligible, which is satisfied in the parameter range we consider
as shown in Figs.~\ref{fig:hd_anly} and \ref{fig:hd_anly4},
We also carried out simulations with $\acc=10^{-3}$ 
and $\Dz (= \Dr)$ in the range of $[10^{-7},10^{-3}]$
to find that the contour map is almost identical, 
confirming that the results are scaled by $\Dz/\acc (= \Dr/\acc)$.

Using the formulas, we derive the boundary of the runaway pile-up 
on the plane of $\Dz/\acc$ $(=\Dr/\acc)$ and $\fpg$. 
Equation~(\ref{eq:rho_anly_diff5})
shows that the boundary is given by $F_{\rm d,net} \simeq h_{\rm d/g}$. 
As shown in Figure~\ref{fig:map}a and b, the boundary
is located at $\Dz (=\Dr) < 3\times 10^{-4}$ for $\fpg < 1$.
In this parameter range, Figure~\ref{fig:hd_anly} shows $h_{\rm d/g} \simeq h_{\rm d/g,*}$.
Since $\Dr /C_{r, \rm diff} \, \acc \ll 1$ for $\acc \sim 10^{-2}$,
$F_{\rm d,net} \simeq f_{\rm d/p} \fpg$.
Therefore the boundary is approximately given by
\begin{align}
\fpg & \simeq f_{\rm d/p}^{-1} h_{\rm d/g,*} \nonumber\\
       & \simeq f_{\rm d/p}^{-1} \left( \frac{2}{3} \frac{\Dz}{\acc} \frac{\Delta x_{\rm subl}/\hg}{\hg/r} \right)^{1/2}  
       \left[ 1 + \frac{2}{3} \frac{\Dr}{\acc} \frac{1}{(\hg/r)(\Delta x_{\rm subl}/\hg)} \right].
\label{eq:boundary}
\end{align}
Adopting $\hg/r=0.04$ and $f_{\rm d/p} = 1/2$ as nominal values,
\begin{eqnarray}
\fpg & \simeq & 0.13 \left(\frac{f_{\rm d/p}}{1/2}\right)^{-1} 
 \left(\frac{\Dz/\acc}{10^{-2}}\right)^{1/2} 
\left(\frac{\hg/r}{0.04}\right)^{1/2}
    \left( \frac{\Delta x_{\rm subl}}{0.1\,\hg} \right)^{1/2} \nonumber \\
  &  & \times  \left[ 1 + \frac{5}{3} \left(\frac{\Dr/\acc}{10^{-2}}\right) 
 \left(\frac{\hg/r}{0.04}\right)^{-1} \left(\frac{\Delta x_{\rm subl}}{0.1\,\hg}\right)^{-1}  \right] \nonumber \\
  & \propto & \left\{
\begin{array}{ll}
\displaystyle
\left( \frac{\Dz}{\acc} \right)^{1/2} \left( \frac{\Dr}{\acc} \right) \Delta x_{\rm subl}^{-1} & \displaystyle [\frac{\Dz}{\acc} > 0.6 \times 10^{-2} ], \label{eq:boundary1}\\
\displaystyle
\left( \frac{\Dz}{\acc} \right)^{1/2} \Delta x_{\rm subl}^{1/2} & \displaystyle [\frac{\Dz}{\acc} < 0.6 \times 10^{-2} ].  \label{eq:boundary2}
\end{array}
\right.  
\end{eqnarray}
This equation explicitly shows that the pile-up condition is
scaled by $\Dr/\acc$ and $\Dz/\acc$.

In Eq.~(\ref{eq:boundary2}), the sublimation width, $\Delta x_{\rm subl}$, is a free parameter.
We have used $\Delta x=0.1 \, \hg$ as a nominal parameter.
The sublimation width is determined by
the sublimation and drift rates of icy pebbles.
The sublimation rate is given by
the partial pressure of water vapor and the disk temperature.
The water vapor partial pressure is affected by radial diffusion.
In Paper II, we numerically solve
the sublimation width taking account of the icy pebble size evolution
and the water vapor partial pressure in a turbulent accretion disk.
Fitting the numerical results, 
the sublimation width for $\Dz=\Dr$ and $\acc=10^{-2}$ is roughly approximated as
\begin{align}
\log_{10} \left(\frac{\Delta x_{\rm subl}}{\hg}\right) 
& \simeq \frac{X_+ - X_-}{2} 
{\rm erf} 
\left[ 3 \log_{10} \left(\frac{\Dr/\acc}{0.08}\right) \right] 
+ \frac{X_+ + X_-}{2}, \nonumber\\
 & X_- = \log_{10} (\Delta x_{\rm subl}/\hg)_{-} = \log_{10} 2; \\
 & X_+ = \log_{10} (\Delta x_{\rm subl}/\hg)_{+} = \log_{10} 0.1,
\label{eq:dx0a2}
\end{align} 
where $(\Delta x_{\rm subl}/\hg)_{-} = 2$ and 
$(\Delta x_{\rm subl}/\hg)_+ = 0.1$ are
the lower $\Dr/\acc$ limit where advection dominates
and higher one where radial diffusion dominats, respectively (Paper II).
The error function smoothly connects the two limits. 
Figures~\ref{fig:map} show 
that the threshold value of $\fpg$ is several times higher than in Fig.~\ref{fig:map}a
for $\Dz/\acc\sim 10^{-4}$,
while it is similar for $\Dz/\acc\sim 10^{-2}$.
In the former case, $\fpg \propto \Delta x_{\rm subl}^{1/2}$ (Eq.~{\ref{eq:boundary2}),
so that $\fpg$ must be
larger in Fig.~\ref{fig:map}b  than in Fig.~\ref{fig:map}a by a factor of $(2/0.1)^{1/2} \sim 4.5$ (Eq.~(\ref{eq:dx0a2})),
while the dependence on $\Delta x_{\rm subl}$ must be slightly weaker 
in the latter case (Eq.~(\ref{eq:dx0a2})).
Thereby, the silicate particle runaway pile-up condition 
with Eq.~(\ref{eq:dx0a2}) is roughly given by
\begin{equation}
\fpg \ga \left(\frac{\Dz/\acc}{3 \times 10^{-2}}\right)^{1/2},
\label{eq:pileup}
\end{equation}
which agrees with the results in Fig.~\ref{fig:map}.

Figure~\ref{fig:map} show the results with $\Dr = \Dz$.
MHD simulations showed $\Dr < \Dz$ \citep{Zhu2015,Yang2018}.
We also performed simulations with $\Dr = 0.1\, \Dz$.
As we can predict from Eq.~(\ref{eq:boundary1}),
the runaway pile-up region is extended to a several times larger value of
$\Dz/\acc$ at $\fpg \sim 1$ in the case of $\Delta x_{\rm subl}=0.1\,\hg$, 
while the runaway pile-up region
is almost the same for smaller values of $\fpg$.
Because $\Dz/\acc$ is very uncertain, we surveyed a rather broad parameter space
and leave the estimate of $\Dz/\acc$ for future study.

\subsubsection{Comparison with \citet{Hyodo2019}}

\citet{Hyodo2019} performed 1D advection-diffusion grid code simulations with
$\acc=10^{-2}$ and $\Dr=\Dz=10^{-3}, 3\times 10^{-3}, 10^{-2}$
to find that the silicate runaway pile-up occurs for $\Dz=\Dr\approx 10^{-1}\acc$ and $\fpg \ge 0.3$
in the case of $K=1$. 
However, Figure~\ref{fig:map}b shows that the silicate runaway pile-up region is restricted to a smaller parameter range, i.e., to values of 
$\Dz/\acc=\Dr/\acc$ which are one order of magnitude smaller for a given $\fpg$ value. 
As already mentioned, this is because \citet{Hyodo2019} assumed that 
 the silicate scale height is $\sim \Delta z_{\rm subl}$ at the snow line and
 it is gradually increased by the vertical turbulent stirring as the particles drift inward whereas 
the Monte Carlo simulations performed here show that the silicate scale height is larger,
due to the effect of a coupled radial and vertical diffusion in a finite-size sublimation width.
As a result, smaller $\Dz$ and $\Dr$ are required for a runway pile-up.

\section{Conclusion and Discussions}

The runaway pile-up of silicate particles released from sublimating icy pebbles
that pass through the snow line is a potential mechanism to form rock-rich planetesimals.
\citet{Ida_Guillot2016} showed that the back-reaction (inertia) of silicate particles to gas drag can lead to a runaway pile-up of dust particles. They provided a simple criterion for this to occur, as a function of both the pebble-to-gas mass-flux ratio and the silicate particle scale-height. 
\citet{Schoonenberg2017} found instead that a runaway pile-up of ice-rich pebbles would be possible and that of silicate particles does not occur. However they did not include the back-reaction of dust particles, and thus found no pile-up of dust particles.  \citet{Hyodo2019} performed 1D diffusion-advection grid code simulations
including turbulent diffusion and the back-reaction to radial drift and diffusion for both icy pebble and silicate particles. They allowed the local turbulent diffusion governing radial and vertical diffusion to differ from the turbulent viscosity controlling the gas mass flux in the disk. They found that a runaway pile-up of either dust or pebbles could be achieved, depending on the values of turbulent diffusion and of the pebble-to-gas mass-flux ratio. However, they had to approximate the calculation of the silicate particle scale-height based on simple arguments. In this work, we revisited this issue. 

We have developed a new 2D ($r$-$z$) Monte Carlo code to simulate
the pile-up of small silicate particles released from sublimating pebbles
in a turbulent protoplanetary disk, taking account of the back-reactions to 
the drift velocity and the diffusion of silicate particles.
From the simulation results, we have derived
semi-analytical formulas for the maximum silicate-to-gas density ratio near the injection region
and the silicate scale height there as a function of the pebble mass flux, $\acc$, $\Dr$, and $\Dz$.
Using the derived formulas, we determined the detailed condition for
the silicate runaway pile-up.
We found that the silicate particle scale height is larger than the \citet{Hyodo2019}'s estimate
due to the coupled effect of radial and vertical diffusion,
and that for the silicate particle runaway pile-up to occur,
$\Dz/\acc \la 10^{-2} \times$ (pebble-to-gas mass flux)$^2$ is required, which is more restrictive 
than \citet{Hyodo2019}'s result ($\Dz/\acc \la 10^{-1}$ and $\fpg \ga 0.3$). 
To clarify if the condition is actually satisfied, detailed non-ideal MHD simulations  are needed
to evaluate $\Dz$, $\Dr$, and $\acc$.  
 
Thus far, we have not included the pile-up of icy pebbles. This would occur upstream, i.e., beyond the snow line, and could thus suppress the supply of pebbles and thus dust grains, affecting the pile-up of dust-rich planetesimals. It is also important to understand when and where runaway pile-ups of icy pebbles and silicate particles may occur in the course of disk evolution.
We will investigate these issues in Paper II. 

One would want to extend the arguments developed in this work to other ice lines, such as for NH$_3$ and CO$_2$.
The pile-up process proposed here occurs only if the Stokes numbers of the particles before and
after the sublimation of a volatile component ($\tau_{\rm s,0}$ and $\tau_{\rm s,1}$, respectively) 
satisfy $\tau_{\rm s,0} \gg \acc \ga \tau_{\rm s,1}$.
It is unlikely that this condition is satisfied at the ice lines of volatile elements other than H$_2$O.
This pile-up process is therefore expected to be effective only for the water snow line.

\begin{acknowledgements}
We thank 
Vardan Elbakyan for providing his simulation data and 
Chao-Chin Yang for helpful comments.
SI was supported by MEXT Kakenhi 18H05438. TG was partially supported by a JSPS Long Term Fellowship at the University of Tokyo.
RH was supported by JSPS Kakenhi JP17J01269 and 18K13600). RH also acknowledges JAXA’s International Top Young program.
SO was supported by JSPS Kakenhi 19K03926 and 20H01948.
ANY was supported by NASA Astrophysics Theory Grant NNX17AK59G and NSF grant AST-1616929.
\end{acknowledgements}


\appendix
\section{The effects of back-reaction to gas motion} 

The silicate dust particle and gas radial velocities taking account of the inertia of the particles 
(``back-reaction'') at $|z| < \hd$ 
are given respectively by \citep{Schoonenberg2017}
\begin{eqnarray}
\vr & = & -\Lambda^2 \frac{2\st}{1+\Lambda^2 \st^2}\eta \vk
+ \Lambda \frac{1}{1+\Lambda^2 \st^2}u_\nu, \label{eq:vr2} \\
u_r & = &
Z \Lambda^2 \frac{2\st}{1+\Lambda^2 \st^2}\eta \vk
+  \Lambda \frac{1 + \Lambda \st^2}{1+\Lambda^2 \st^2} u_\nu,
 \label{eq:ur2}
\end{eqnarray}
where $\st$ is Stokes number of the particles, $Z= \rhod/\rhog$, $\Lambda = \rhog/(\rhog + \rhod) =1/(1+Z)$, $u_\nu $ is an unperturbed disk gas accretion velocity given by
\begin{eqnarray}
u_\nu \simeq - \frac{3\nu}{2r} 
          \simeq  -\frac{3\acc}{2C_{\eta}}  \,\eta \vk,
\end{eqnarray}
and $C_{\eta}= \eta/(H_{\rm g}/r)^2$.
When we consider the icy pebbles, the subscript ``d'' is replaced by ``p.''  
Equations~(\ref{eq:vr2}) and (\ref{eq:ur2}) are rewritten as
\begin{align}
\vr & \simeq -\frac{\Lambda}{1+\Lambda^2 \st^2}
\left( 2 \Lambda \,\st  + \frac{3}{2C_{\eta}}  \acc \right)  \eta \vk, \label{eq:vr3} \\
u_r & \simeq -\frac{\Lambda}{1+\Lambda^2 \st^2}
\left( - 2 Z \Lambda \,\st  + \frac{3}{2C_{\eta}}(1 + \Lambda \st^2) \acc \right)  \eta \vk.
 \label{eq:ur3}
\end{align}

We adopt a two-layer model: a dust-rich midplane layer 
with scale height $H_{\rm d}$ and a dust-poor upper layer.
To evaluate gas surface density, we use the vertically averaged $u_r$, such that
\begin{eqnarray}
Z_{\Sigma} = \frac{\sigd}{\sigg} 
\simeq \frac{\bar{u}_r}{\vr} \frac{f_{\rm d/p} \mdotp}{\mdotg} 
= \frac{\bar{u}_r}{\vr} f_{\rm d/p} \fpg,
\label{eq:Z2}
\end{eqnarray}
where 
\begin{align}
\bar{u}_r & \simeq \frac{u_r|_{z=0} \, H_{\rm d} + u_\nu (H_{\rm g} - H_{\rm d})}{H_{\rm g}}
= \hdg \, u_r|_{z=0} + (1 - \hdg) u_\nu. 
\label{eq:bar_u}
\end{align}
Substituting Eq.~(\ref{eq:ur3}) into this equation, we obtain
\begin{align}
\bar{u}_r & \simeq \frac{\Lambda}{1+\Lambda^2 \st^2} \nonumber\\
 & \times \left[
\left(\frac{2\st}{1+Z} + \frac{3\acc}{2C_{\eta}} \right)
Z \, \hdg
- \frac{3\acc}{2C_{\eta}} (1 + Z + \Lambda \st^2) \right] \eta \vk.
\label{eq:bar_u2}
\end{align}

We note that
even the vertically averaged gas motion is outward for
\begin{equation}
\frac{\acc}{\st} 
< \frac{4C_\eta}{3} \hdg \frac{Z}{(1+Z)[1+Z(1-\hdg)]}, 
\label{eq:outward1}
\end{equation}
where we assumed $\st^2 \ll 1$.
For $Z \la 1$ and $C_\eta = 11/8$, this condition is reduced to 
\begin{equation}
\frac{\acc}{\st} \la \frac{11}{6}\hdg \,Z.
\label{eq:outward2}
\end{equation}
In the case of silicate dust particles, 
since $\st \ll \acc$, this condition is not satisfied and the gas flow is always inward.

For pebbles, the outward flow condition
is the same as Eq.~(\ref{eq:outward2}) with ``d'' replaced by ``p''.
The condition is less restrictive than that for silicate particles,
however, it is not satisfied within the parameter range in this paper,
as shown below.
We usually consider the cases of $\stpeb > \acc$, and 
$\hpg \simeq (\Dz/\stpeb)^{1/2}$.
In this case, the modified 
Eq.~(\ref{eq:outward2}) is $Z \ga (6/11)(\acc/\Dz) \hpg$.
If this equation is satisfied, the (vertically averaged) gas flow is outward beyond the snow line.
However, in our case, $Z$, $\acc$, and $\Dz$ are not independent.
Equation~(\ref{eq:Zice}) is $Z \simeq (1/2)(3/4C_\eta)(\Dz/\acc)\hpg\fpg$ for $Z<1$.
Thus, the outward flow condition in the icy pebble region is $\fpg \ga 2(\acc/\Dz)^2$.
It is out of the parameter range that we cover in this paper.
In the parameter regimes we cover, 
$\bar{u}_r$ is always negative (inward flow) in the regions of
both silicate particles and icy pebbles.

Substituting Eqs.~(\ref{eq:bar_u2}) and (\ref{eq:vr2}) into Eq.~(\ref{eq:Z2}),
 \begin{align}
Z \, \hdg & \simeq Z_\Sigma \simeq \frac{\bar{u}_r}{\vr} 
f_{\rm d/p} \fpg \\
 &  \simeq 
     \frac{-2 \frac{Z}{1+Z} \, \hdg \st + \frac{3}{2C_{\eta}}\acc \left[1 + Z (1-\hdg)
      + \frac{\st^2}{1+Z}\right]}{2 \Lambda \,\st  + \frac{3}{2C_{\eta}} \acc} f_{\rm d/p} \fpg. 
\label{eq:Zdust}
\end{align}
Now we adopt the approximation only appropriate for the silicate dust,
$\st \ll \acc \ll 1$.
With this approximation, 
Eq.~(\ref{eq:Zdust}) is reduced to
\begin{align}
Z \, \hdg & \simeq 
\left[1 + Z (1-h_{\rm d/g}) \right] f_{\rm d/p} \fpg,
\end{align}
which is solved in terms of $Z$ as
\begin{align}
Z & \simeq \frac{f_{\rm d/p} \fpg}{\hdg - (1-\hdg) f_{\rm d/p} \fpg}.
\label{eq:Zdust2}
\end{align}
As we discussed in Section 2, 
the pile-up of silicate particles is radially local and
the corresponding local surface density variation of $\sigg$ could be smoothed out.
In that case, it may be better to use $u_r = u_\nu$ than $\bar{u}_r$ here,
and we obtain 
\begin{align}
Z & \simeq \frac{f_{\rm d/p} \fpg}{\hdg - f_{\rm d/p} \fpg}.
\label{eq:Zdust3}
\end{align}
Because $\hdg \ll 1$ for the pile-up case, the pileup condition differs only slightly
between Eqs.~(\ref{eq:Zdust2}) and (\ref{eq:Zdust3}). 

From Eqs.~(\ref{eq:Zdust2}) and (\ref{eq:Zdust3}), 
the runaway pile-up conditions of silicate particles 
are given by
\begin{align}
\fpg & > \frac{f_{\rm d/p}^{-1}}{1-\hdg}\hdg \:{\rm or} \: > f_{\rm d/p}^{-1} \hdg & [\rm silicate \: particles],\label{eq:pileup_sil}
\end{align}
This equation implies that the particle scale height is the key parameter
for the runaway pile-up:
the pile-up is favored for a smaller scale height.
In Section 4.3, we discuss the effect of radial and vertical diffusion on the 
silicate particle scale height $\hdg$, because it is not in an equilibrium state.

On the other hand, for pebbles, $\acc \ll \stpeb \ll 1$ is an appropriate approximation.
From $\acc \ll \stpeb$, we can also assume $\hpg \ll 1$.
With these approximations, Eq.~(\ref{eq:Zdust}) with 
``d'' replaced by ``p'' and $f_{\rm d/p}$ by 1 is reduced to
\begin{align}
Z \, \hpg &  \simeq 
\frac{3}{4C_{\eta}}
\frac{\acc}{\stpeb}(1+Z)^2 \fpg.
\label{eq:Zice}
\end{align}
This quadratic equation does not have a solution if
\begin{align}
\left( 1 - \frac{2C_{\eta}}{3}
\frac{\stpeb}{\acc} \frac{\hpg}{\fpg} \right)^2 - 1 < 0.
\end{align}
Therefore, the runaway pile-up condition for pebbles is
\begin{align}
\fpg & > \frac{C_{\eta}}{3}
\frac{\stpeb}{\acc} \hpg.
\label{eq:pebble_runaway}
\end{align}
We will discuss the pebble runaway
pile-up condition with the effect of the vertical stirring by Kelvin-Hermholtz Instability
in more details in Paper II.

\section{Consequences of outward diffusion over the snow line}

In Eq.~(\ref{eq:IG3}), the radial diffusion effect of silicate particles is not included.
The (scaled) radial diffusion flux of silicate particles is
\begin{equation}
F_{\rm d,D} \sim 2\pi r \, \Dr \hg^2 \Omega \frac{d \sigd}{d r} \dot{M}^{-1}_{\rm g}
\sim 2\pi \, C_F \, r \, \Dr \hg^2 \Omega \frac{\sigd}{\hg} \dot{M}^{-1}_{\rm g},
\label{eq:FD}
\end{equation}
where $C_F$ is an unknown numerical factor.
We calibrate $C_F$ by the results of Monte Carlo simulations
in Figs.~\ref{fig:hd_anly} and \ref{fig:hd_anly4} as follows.
The (scaled) net silicate dust mass flux, $F_{\rm d,net} \sim f_{\rm d/p} F_{\rm p/g} - F_{\rm d,D}$,
in a steady state should satisfy
$F_{\rm d,net} \sim 3\pi \sigd \acc \hg^2 \Omega \dot{M}^{-1}_{\rm g}$. 
Substituting Eq.~(\ref{eq:FD}) into this equation, we obtain
\begin{align}
F_{\rm d,D} & \sim f_{\rm d/p} {F}_{\rm p/g}\times \left[1+ \frac{3}{2\,C_F}\frac{\hg}{r}\frac{\acc}{\Dr}\right]^{-1}.
 \end{align} 
Accordingly, $(\rhod/\rhog)_0$ is
given by Eq.~(\ref{eq:IG3}) with $f_{\rm d/p} F_{\rm p/g}$ replaced by  
$F_{\rm d,net} = f_{\rm d/p} F_{\rm p/g} - F_{\rm d,D}$, as
\begin{eqnarray}
\left(\frac{\rho_{\rm d}}{\rho_{\rm g}}\right)_0 =
\frac{F_{\rm d,net}}{h_{\rm d/g}  - F_{\rm d,net}},
\label{eq:rho_anly_diff}
\end{eqnarray}
where 
\begin{equation}
F_{\rm d,net} = f_{\rm d/p} F_{\rm p/g}
\left\{1 - \left[1+ \frac{3}{2 \, C_F}\frac{\hg}{r}\frac{\acc}{\Dr}\right]^{-1} \right\}.
\label{eq:rho_anly_diff2}
\end{equation}
Figures~\ref{fig:hd_anly} and  \ref{fig:hd_anly4} 
show that the analytical results by Eq.~(\ref{eq:rho_anly_diff}) with $C_F = 0.2$ 
and they fit the numerical results as long as $\rhod/\rhog \la 1$.
If the effect of sticking to icy pebbles
by silicate particles that diffuse beyond the snow line is considered,
most of the silicate grains beyond the snow line
may eventually come back with the pebbles,
so that $F_{\rm d,net}$ would become more similar to $f_{\rm d/p} F_{\rm p/g}$.

\end{document}